\documentclass[12pt]{article}
\usepackage{epsfig}
\usepackage{graphicx}
\usepackage{amsmath}

\begin{document}

 \newcommand{\beq}{\begin{equation}}
\newcommand{\eeq}{\end{equation}}
\newcommand{\bea}{\begin{eqnarray}}
\newcommand{\eea}{\end{eqnarray}}
\newcommand{\beqn}{\begin{eqnarray}}
\newcommand{\eeqn}{\end{eqnarray}}
\newcommand{\beas}{\begin{eqnarray*}}
\newcommand{\eeas}{\end{eqnarray*}}
\newcommand{\defi}{\stackrel{\rm def}{=}}
\newcommand{\non}{\nonumber}
\newcommand{\bquo}{\begin{quote}}
\newcommand{\enqu}{\end{quote}}
\newcommand{\qt}{\tilde q}
\newcommand{\m}{\tilde m}
\newcommand{\trho}{\tilde{\rho}}
\newcommand{\tn}{\tilde{n}}
\newcommand{\tN}{\tilde N}

\newcommand{\gsim}{\lower.7ex\hbox{$
\;\stackrel{\textstyle>}{\sim}\;$}}
\newcommand{\lsim}{\lower.7ex\hbox{$
\;\stackrel{\textstyle<}{\sim}\;$}}


\def\de{\partial}
\def\Tr{ \hbox{\rm Tr}}
\def\const{\hbox {\rm const.}}
\def\o{\over}
\def\im{\hbox{\rm Im}}
\def\re{\hbox{\rm Re}}
\def\bra{\langle}\def\ket{\rangle}
\def\Arg{\hbox {\rm Arg}}
\def\Re{\hbox {\rm Re}}
\def\Im{\hbox {\rm Im}}
\def\diag{\hbox{\rm diag}}


\def\QATOPD#1#2#3#4{{#3 \atopwithdelims#1#2 #4}}
\def\stackunder#1#2{\mathrel{\mathop{#2}\limits_{#1}}}
\def\stackreb#1#2{\mathrel{\mathop{#2}\limits_{#1}}}
\def\Tr{{\rm Tr}}
\def\res{{\rm res}}
\def\Bf#1{\mbox{\boldmath $#1$}}
\def\balpha{{\Bf\alpha}}
\def\bbeta{{\Bf\beta}}
\def\bgamma{{\Bf\gamma}}
\def\bnu{{\Bf\nu}}
\def\bmu{{\Bf\mu}}
\def\bphi{{\Bf\phi}}
\def\bPhi{{\Bf\Phi}}
\def\bomega{{\Bf\omega}}
\def\blambda{{\Bf\lambda}}
\def\brho{{\Bf\rho}}
\def\bsigma{{\bfit\sigma}}
\def\bxi{{\Bf\xi}}
\def\bbeta{{\Bf\eta}}
\def\d{\partial}
\def\der#1#2{\frac{\d{#1}}{\d{#2}}}
\def\Im{{\rm Im}}
\def\Re{{\rm Re}}
\def\rank{{\rm rank}}
\def\diag{{\rm diag}}
\def\2{{1\over 2}}
\def\ntwo{${\mathcal N}=2\;$}
\def\nfour{${\mathcal N}=4\;$}
\def\none{${\mathcal N}=1\;$}
\def\ntwot{${\mathcal N}=(2,2)\;$}
\def\ntwoo{${\mathcal N}=(0,2)\;$}
\def\x{\stackrel{\otimes}{,}}

\def\ba{\beq\new\begin{array}{c}}
\def\ea{\end{array}\eeq}
\def\be{\ba}
\def\ee{\ea}
\def\stackreb#1#2{\mathrel{\mathop{#2}\limits_{#1}}}

\def\Tr{{\rm Tr}}
\newcommand{\vp}{\varphi}
\newcommand{\pt}{\partial}
\newcommand{\ve}{\varepsilon}
\renewcommand{\theequation}{\thesection.\arabic{equation}}

\setcounter{footnote}0

\vfill

\begin{titlepage}

\begin{flushright}
FTPI-MINN-09/13, UMN-TH-2741/09\\
April 6, 2009
\end{flushright}

\begin{center}
{  \Large \bf   Non-Abelian Duality  and Confinement in
 \boldmath{\ntwo} 
 Supersymmetric QCD}

\vspace{4mm}

 {\large
 \bf    M.~Shifman$^{\,a}$ and \bf A.~Yung$^{\,\,a,b}$}
\end {center}

\begin{center}


$^a${\it  William I. Fine Theoretical Physics Institute,
University of Minnesota,
Minneapolis, MN 55455, USA}\\
$^{b}${\it Petersburg Nuclear Physics Institute, Gatchina, St. Petersburg
188300, Russia
}
\end{center}

\begin{center}
{\large\bf Abstract}
\end{center}

In \ntwo supersymmetric QCD with the U($N$) gauge group
and  $N_f>N $ we study the crossover transition
from the weak coupling regime at large $\xi$ to  strong coupling  
at small $\xi$ where $\xi$ is the Fayet--Iliopoulos parameter.
We find that at strong coupling a dual non-Abelian weakly coupled ${\mathcal N}=2$
theory exists which describes low-energy physics at small $\xi$.
The dual gauge group is U$(N_f-N)$ and the dual theory has $N_f$ flavors
of light dyons, to be compared with $N_f$ quarks in the original U($N$)
theory. Both, the original and dual theories are Higgsed and share the same global symmetry
${\rm SU}(N)\times  {\rm SU}(N_f-N) \times {\rm U}(1)$, albeit
the physical meaning of the ${\rm SU}(N)$ and $ {\rm SU}(N_f-N)$ factors
is different in the large- and small-$\xi$ regimes.
Both regimes support non-Abelian semilocal strings.
 In each of these two regimes
particles that are in the adjoint representations with respect to one of the factor groups
exist in two varieties: elementary fields and composite states bound by strings.
These varieties interchange upon transition from one regime to the other. 
We conjecture that the composite stringy states can be related to Seiberg's $M$ fields. 
 The bulk duality that we observed translates into a two-dimensional
duality on the world sheet of the non-Abelian strings. At large $\xi$ the 
internal dynamics of the semilocal non-Abelian strings is described by
the sigma model of $N$ orientational and $(N_f-N)$ size moduli, while
at small $\xi$ the roles of orientational and size moduli interchange.
The BPS spectra of two dual sigma models (describing confined 
monopoles/dyons of the bulk theory) coincide. It would be interesting to trace parallels
between the non-Abelian duality we found and string theory constructions.

\vspace{2cm}

\end{titlepage}

\newpage

\tableofcontents

\newpage

\section {Introduction}
\label{intro}
\setcounter{equation}{0}

In this paper  we continue studying  transitions from weak to strong coupling
in \ntwo supersymmetric QCD induced by a change of parameters. The investigation began
in \cite{SYcross} where we considered Yang--Mills theory with the gauge group U$(N)$
and $N$ matter hypermultiplets in the fundamental representation. The adjustable parameters in this theory are
the Fayet--Iliopoulos (FI) \cite{FayIl} coefficient $\xi$ and the quark mass differences described by a set of parameters
 $\Delta m_{AB}$. The overall scale is set by a dynamical parameter $\Lambda$. We started from $\xi\gg\Lambda$
while our task was to penetrate in the domain $\xi\lsim\Lambda$ and small (vanishing) $\Delta m_{AB}$.
In the former limit the theory is weakly coupled and one can obtain a reliable quasiclassical description of physics
directly from the given microscopic theory. In particular, at $\Delta m_{AB}=0$ there emerge non-Abelian strings
\cite{HT1,ABEKY,SYmon}
whose world-sheet dynamics is described by supersymmetric CP$(N-1)$ model (for reviews see \cite{SYrev,Trev,Jrev,Trev2}).
These strings confine monopoles \cite{SYmon,HT2}. Nonperturbative  light ``mesonic" states are monopole-antimonopole pairs
connected by two non-Abelian strings.

On the other hand, at $\xi\lsim\Lambda$ our microscopic theory is strongly coupled.
To develop an effective low-energy description of physics in this domain of small $\xi$  
(and small $|\Delta m_{AB}|$) we had to derive
a dual weakly coupled theory. The dual theory turned out to be Abelian, based on U(1)$^{N-1}$.
Moreover, we found that the light matter sector in this Abelian theory consisted of certain dyons, which condense in the vacuum resulting in Abelian strings of the Abrikosov--Nielsen--Olesen (ANO) \cite{ANO} type.
The  light ``mesonic" states built from the monopole-antimonopole pairs
connected by two strings survive, albeit these strings are totally different from those
in the large-$\xi$ small-$|\Delta m_{AB}|$ domain. We came to the conclusion that the transition from the non-Abelian
to Abelian (low-energy) regimes was of a crossover type rather than a phase transition.\footnote{It is worth adding that
it does become a phase transition at $N=\infty$.}

\begin{figure}
\epsfxsize=5cm
\centerline{\epsfbox{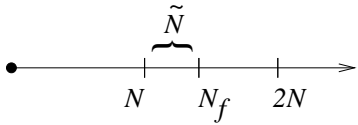}}
\caption{\small The number of flavors exceeds the number of colors,
$\tilde N\equiv N_f - N >0.$}
\label{voenn}
\end{figure}

\begin{figure}
\epsfxsize=7cm
\centerline{\epsfbox{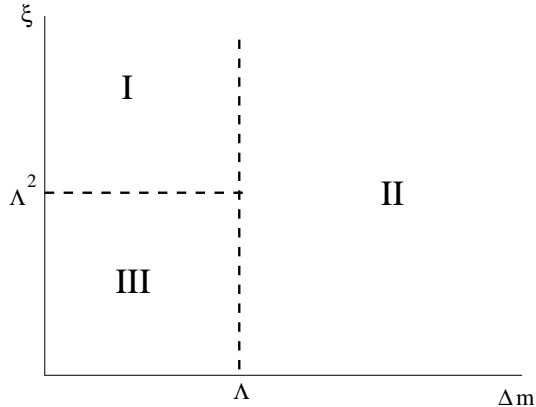}}
\caption{\small Various regimes in  \ntwo QCD
  are  separated by crossovers.    The dynamical scale of our microscopic
  non-Abelian gauge theory  is
 represented by the  parameter $\Lambda$. }
\label{figphasediag}
\end{figure}

In this paper we extend the scope of our studies to cover the case of a larger number of the 
fundamental matter hypermultiplets,
i.e.  $N_f > N$, see Fig. \ref{voenn}. Other than that, the microscopic theory we work with is the same as in
\cite{SYcross}. Namely, we deal with  \ntwo supersymmetric QCD with the  gauge group U$(N)$ and 
the Fayet--Iliopoulos term. Although 
$N_f>N$, we limit ourselves to $N_f<2N$ to keep asymptotic freedom in our microscopic theory.
The Fayet--Iliopoulos  term $\xi\neq 0$
triggers condensation of $N$ squark fields. The parameter space of this theory includes the FI parameter 
$\xi$ and the squark mass differences 
\beq
\Delta m_{AB}=m_A-m_B\,,\qquad A,B=1,...,N_f\,.
\label{sqmd}
\eeq
Various regimes  of the theory in the $\{ \xi,\,\Delta m\}$ plane are schematically shown in Fig.~\ref{figphasediag}. 
The vertical axis in this figure denotes the values of the FI parameter $\xi$ while the horizontal
axis  schematically represents all quark mass differences.

At $\xi\gg \Lambda^2 $ the theory is at weak 
coupling. Perturbative and nonperturbative spectra, and all interactions
can be exhaustively analyzed using quasiclassical methods.   
 In  the limit of 
degenerate quark masses $\Delta m_{AB}=0$
the microscopic  theory at hand  has an unbroken global SU$(N)$ symmetry
which is a diagonal combination of SU$(N)_{\rm color}$
and an SU$(N)$ subgroup of the flavor SU$(N_f)$ group acting in the theory.
Thus, the
color-flavor locking takes place,
see Sect.~\ref{bulk}.  All light states come in the adjoint and singlet 
representations of the unbroken SU$(N)_{\rm diag}$. 

Much in the same way as in \cite{SYcross} the  theory with $N_f >N $  supports
non-Abelian flux tubes (strings) in the weak coupling domain I. In fact, at $N_f>N$ these strings are semilocal
(for a review on Abelian semilocal strings see e.g. \cite{AchVas}).
Internal dynamics of  semilocal non-Abelian strings is described by
 two-dimensional \ntwo 
supersymmetric sigma model with toric target space \cite{HT1,HT2,SYsem,Jsem,Jsemp}. It contains 
 $N$ orientational and $\tN$ size moduli, where
\beq
\tilde{N}\equiv N_f-N.
\label{tilen}
\eeq
Since the squark fields are condensed in the domain I and the theory is fully   Higgsed,
the monopoles are attached to strings. In fact,
in the U$(N)$ gauge theory the monopoles of the SU$(N)$ sector represent junctions of two distinct degenerate strings
and are seen as kinks in the world-sheet sigma model on the non-Abelian string \cite{T,SYmon,HT2},
see also the review \cite{SYrev}.

The domain II is that of the {\em Abelian} Higgs regime at weak coupling. As we increase  
  $\Delta m_{AB}$, the (off diagonal) $W$ bosons and their superpartners become exceedingly 
heavier and decouple from the low-energy spectrum. We are  left with the
photon (diagonal) gauge fields and their quark \ntwo superpartners. Explicit
breaking of the flavor symmetry by $\Delta m_{AB}\neq 0$ leads to the loss
of non-Abelian nature of the string solutions; they  become
Abelian (the so-called $Z_N$) strings. 

Finally, as we reduce $\xi$ and  $|\Delta m_{AB}|$ below $\Lambda$, we enter the strong coupling 
domain III. Because of strong coupling the original microscopic theory is not 
directly analytically tractable here. Our task is to find a weakly coupled dual theory
which describes physics in this domain. We show that at $N_f>N$
such a dual theory does exist and, moreover, it is non-Abelian, with 
 the dual gauge group
\beq
{\rm U}(\tN)\times {\rm U}(1)^{N-\tN}\,,
\label{dualgaugegroup}
\eeq
and $N_f$ flavors of  charged non-Abelian dyons. The quarks we started from in the domain I 
transform themselves into dyons
due to monodromies as we reduce $|\Delta m_{AB}|$. In its gross features the dual \ntwo theory
we found is
similar to Seiberg's dual \cite{Seiberg:1994pq} (for reviews see \cite{Shifintri}) to our original microscopic theory. 
Because
$N_f>2\tN$,  the dual theory is infrared (IR) free rather than
asymptotically free. This result is in perfect match with the
results obtained in \cite{APS} where the
dual non-Abelian gauge group SU$(\tN)$ was identified at the root of a
baryonic branch in the SU$(N)$ gauge theory with massless quarks, see also \cite{CKM}.
In  the limit of 
degenerate quark masses $\Delta m_{AB}=0$ and small $\xi$,
the dual theory has an unbroken global diagonal SU$(\tN)$ symmetry.
It is obtained as a result of the spontaneous breaking of the gauge U$(\tN )$
group and an SU$(\tN )$ subgroup of the flavor SU$(N_f)$ group. 
Thus, the color-flavor locking takes place in the dual theory as well, much in the same way
as in the original microscopic theory in the domain I, albeit the preserved diagonal symmetry is different.
The light states come in adjoint and singlet representations of the global SU$(\tN )$. Thus, the low-energy spectrum of the theory
in the domain III is dramatically different from that of domain I. Excitation spectra are arranged in
different representations of the global unbroken groups, SU$(N)$ and SU$(\tN )$, respectively.
Let us ask ourselves how this can happen in the absence of a phase transition?

 Both, the original and dual theories are Higgsed and share the same global symmetry
$${\rm SU}(N)\times  {\rm SU}(\tN) \times {\rm U}(1)\,.$$
To answer the above question we investigate how all states belonging to the adjoint representation
of either ${\rm SU}(N)$ or ${\rm SU}(\tN)$ evolve when we vary $\xi$ and cross the boundary of the
large- and small-$\xi$ domains. In the large-$\xi$ domain the original theory is at weak coupling
while the dual is at strong coupling and vice versa. It turns out that in  both regimes
we have particles which are adjoint with respect to ${\rm SU}(N)$ and ${\rm SU}(\tN)$.
They come in two varieties: as elementary fields and as composite mesons whose constituents are
bound together by strings. For instance, at small $\xi$ the adjoints in ${\rm SU}(\tN)$
are elementary while the adjoints in ${\rm SU}(N)$ are composite.
At large $\xi$ their roles interchange. The spectrum as a whole is smooth.
The phenomenon of level crossing takes place {\em en route}, at the crossover transition.

Next we show that monopoles are still attached to strings in the domain III at small (but 
nonvanishing) $\xi$. They are represented
by junctions of two different non-Abelian strings of the dual bulk theory and seen as kinks
in the dual world-sheet theory on the string. However,  since in the domain III it is the condensation of dyons
that ensures complete Higgsing of the gauge SU$(\tN)$ group,
we in fact deal with oblique confinement \cite{thooft}.
This result provides a counterexample to a commonly
accepted belief that if monopoles are confined in the original theory, then the quarks 
of the original theory should be confined in the
dual one. We show that monopoles rather than quarks are confined in the domain III.
This observation presumably solves a paradox noted in \cite{Shifman:2007kd}.
Thus, the non-Abelian duality we found is {\em not} the  electro-magnetic duality. This should
be contrasted with the Abelian Seiberg--Witten duality \cite{SW1,SW2}, which {\em is} the electro-magnetic duality.

Three above-mentioned regimes of our microscopic theory --- three domains shown in Fig.~\ref{figphasediag} ---
are arguably separated by crossovers, much in the same way
as it happens in the case $N_f=N$ \cite{SYcross}. In Ref.~\cite{SYcross}
 we argued that the transitions between the domains I, II, and III are crossovers rather than phase transitions.
 Now we will provide further evidence in favor of  crossovers which can 
 be summarized as follows.
 
\vspace{2mm}

(i)  In the equal quark mass limit the domains I and III have Higgs branches of the same dimensions
and the same pattern of global symmetry breaking, see Sect.~\ref{bulkdual}.

\vspace{1mm}

(ii) For generic masses $\Delta m_{AB}\neq 0$  all three regimes have the same number of isolated vacua 
at nonvanishing $\xi$, see Sect.~\ref{split}.

\vspace{1mm}

(iii) Each of these vacua has the same number ($=N$) of different elementary strings in all three
domains. Moreover,
BPS spectra of excitations on the non-Abelian string coincide in the domains I and III,
see Sect.~\ref{2Ddual}.

\vspace{2mm}

Still, as we will show in detail in the bulk of the paper, both the perturbative
spectra and confining strings are dramatically different in the domains I, II, and III. There are certain curves
of marginal stability (CMS) separating these domains. Upon crossing these CMS,
certain elementary particles (like $W$-bosons)
 decay into magnetically charged states. At nonzero $\xi$ these
states are confined and cannot move far apart. They become mesons formed by (anti)monopoles
and  dyons bound together by confining strings.
If, as was claimed above,
 we have a crossover rather than a phase transition between the domains I and III,
then adjoints in the global unbroken symmetry SU$(N)_{C+F}$ 
(present in the domain I) cannot just disappear upon passing in the domain III. Although heavy and invisible in the 
low-energy effective action, they still must survive as particles in the domain III.
We identify these adjoints of SU$(N)_{C+F}$ with composite mesons bound by strings.

Another issue to be discussed in the present paper is a possible
 origin of Seiberg's mesonic fields $M$ \cite{Seiberg:1994pq} which appear
in the dual bulk theory 
when we break \ntwo supersymmetry by the superpotential mass term $\mu{\mathcal A}^2$
  for the adjoint fields and take the limit 
$\mu\to\infty$ thus converting our theory into \none QCD.  The 
composite  mesons formed by (anti)monopoles
and  dyon bound by confining strings are good candidates for 
Seiberg's mesonic fields $M$. While they are heavy in the \ntwo limit, 
they might well become
light in the \none limit. Our arguments in favor of this conjecture
are presented  in Sect.~\ref{LEbulkdual}.

Our results are  in complete parallel
with the situation in the 
special case $N_f=N$  
analyzed in \cite{SYcross}. In this case the domain III
is nothing but the Abelian
Seiberg--Witten confinement \cite{SW1,SW2}. 
The set of  light surviving states includes
photons and dyons with certain quantum numbers. 
The $W$ bosons and their superpartners
decay on the curves of the marginal stability as we move inside III. They are heavy and form
(anti)monopole/dyon stringy mesons at nonzero $\xi$ filling the adjoint representation
of SU$(N)_{C+F}$.

When we speak of dual pairs of theories a clarifying remark is in order. 
There are two slightly different formulations of duality. 
In the first one we start from two different microscopic theories
and show that both theories coincide in the  infrared limit; the infrared description can be strongly coupled, as, say,
in the middle of the conformal window \cite{Seiberg:1994pq}.
In the second formulation, within the given microscopic theory, we identify
two effective theories describing physics at large distances --
one is weakly coupled in a certain domain of parameters where  the other
is strongly coupled and {\em vice versa}. This is the strategy of Seiberg and Witten \cite{SW1}
who, given the SU(2) Yang--Mills theory with ${\mathcal N}=2$, identified a low-energy U(1)
theory and then dualized it to demonstrate the dual Meissner effect
upon a small \ntwo-breaking mass deformation of the original SU(2) theory. 
Our consideration follows the logic of that of \cite{SW1}.

Duality of the bulk theories translates into two-dimensional
duality on the world sheet of the non-Abelian string. The dual SU$(\tN)$ bulk theory 
in the quasiclassical regime supports
non-Abelian semilocal strings. Their internal dynamics is described by two-dimensional \ntwo 
toric sigma model with   $\tN$ orientational and $N$ 
size moduli. Thus, the role of orientational and size moduli interchanges in the domain III as compared with
the domain I. We demonstrate that the BPS spectra of two dual world-sheet theories are
the same.

The general outline of the paper is as follows. In Sect.~\ref{bulk} we review our basic microscopic theory,
and discuss BPS-saturated flux tubes it supports in the domain I.
 We outline the structure of the world-sheet theory on the strings,
which, in the case at hand, is a toric \ntwo sigma model. In Sect.~\ref{bulkdual}
we present a detailed consideration of the transition from the domain I to III. We choose an instructive example $N=3$ and $N_f=5$
and trace the fate of the quarks in their evolution from the domain I to III under a special choice of the quark masses.
Deformations of the quark masses are studied in Sect.~\ref{split}.
In Sect.~\ref{conf} we consider monopoles attached to the strings.
In Section \ref{mmesons} we address evolution and transmutations of the adjoint particles
vs variation of $\xi$ on the way from the domain I to III, a question
which is central for understanding consistency of our picture.
Section \ref{2Ddual} is devoted to the evolution of the world-sheet theory on the way from the domain I to III.
 Section \ref{conclu}
summarizes our conclusions.

\section{Large values of the  FI parameter \\
(Domains I and II)}
\label{bulk}
\setcounter{equation}{0}

In this section  we will briefly review
main features of our basic   theory ---
\ntwo QCD with the  gauge group U($N$) and $N_f$ quark flavors. As shown in Fig.~\ref{voenn}, we assume
$N_f>N$ but $N_f<2N$. The latter inequality ensures  asymptotic freedom of the original microscopic
theory. Then we summarize main features of the  non-Abelian strings in this theory 
\cite{HT1,ABEKY,HT2,SYmon,SYsem,Jsem}.

\subsection{Basic microscopic theory}

The field content is as follows. The \ntwo vector multiplet
consists of the  U(1)
gauge field $A_{\mu}$ and the SU$(N)$  gauge field $A^a_{\mu}$,
where $a=1,..., N^2-1$, and their Weyl fermion superpartners
 plus
complex scalar fields $a$, and $a^a$ and their Weyl superpartners.
The $N_f$ quark multiplets of  the U$(N)$ theory consist
of   the complex scalar fields
$q^{kA}$ and $\tilde{q}_{Ak}$ (squarks) and
their   fermion superpartners, all in the fundamental representation of 
the SU$(N)$ gauge group.
Here $k=1,..., N$ is the color index
while $A$ is the flavor index, $A=1,..., N_f$. We will treat $q^{kA}$ and $\tilde{q}_{Ak}$
as rectangular matrices with $N$ rows and $N_f$ columns.

The bosonic part of our basic 
theory has the form  (for details see  the review paper \cite{SYrev})
\beqn
S&=&\int d^4x \left[\frac1{4g^2_2}
\left(F^{a}_{\mu\nu}\right)^2 +
\frac1{4g^2_1}\left(F_{\mu\nu}\right)^2
+
\frac1{g^2_2}\left|D_{\mu}a^a\right|^2 +\frac1{g^2_1}
\left|\partial_{\mu}a\right|^2 \right.
\nonumber\\[4mm]
&+&\left. \left|\nabla_{\mu}
q^{A}\right|^2 + \left|\nabla_{\mu} \bar{\tilde{q}}^{A}\right|^2
+V(q^A,\tilde{q}_A,a^a,a)\right]\,.
\label{model}
\eeqn
Here $D_{\mu}$ is the covariant derivative in the adjoint representation
of  SU$(N)$, while
\beq
\nabla_\mu=\partial_\mu -\frac{i}{2}\; A_{\mu}
-i A^{a}_{\mu}\, T^a\,.
\label{defnabla}
\eeq
We suppress the color  SU($N$)  indices of the matter fields. The normalization of the 
 SU($N$) generators  $T^a$ is as follows
$$
{\rm Tr}\, (T^a T^b)=\mbox{$\frac{1}{2}$}\, \delta^{ab}\,.
$$
The coupling constants $g_1$ and $g_2$
correspond to the U(1)  and  SU$(N)$  sectors, respectively.
With our conventions, the U(1) charges of the fundamental matter fields
are $\pm1/2$, see Eq.~(\ref{defnabla}).

The scalar potential $V(q^A,\tilde{q}_A,a^a,a)$ in the action (\ref{model})
is the  sum of the $D$ and  $F$  terms,
\beqn
V(q^A,\tilde{q}_A,a^a,a) &=&
 \frac{g^2_2}{2}
\left( \frac{1}{g^2_2}\,  f^{abc} \bar a^b a^c
 +
 \bar{q}_A\,T^a q^A -
\tilde{q}_A T^a\,\bar{\tilde{q}}^A\right)^2
\nonumber\\[3mm]
&+& \frac{g^2_1}{8}
\left(\bar{q}_A q^A - \tilde{q}_A \bar{\tilde{q}}^A -N \xi\right)^2
\nonumber\\[3mm]
&+& 2g^2_2\left| \tilde{q}_A T^a q^A \right|^2+
\frac{g^2_1}{2}\left| \tilde{q}_A q^A  \right|^2
\nonumber\\[3mm]
&+&\frac12\sum_{A=1}^{N_f} \left\{ \left|(a+\sqrt{2}m_A +2T^a a^a)q^A
\right|^2\right.
\nonumber\\[3mm]
&+&\left.
\left|(a+\sqrt{2}m_A +2T^a a^a)\bar{\tilde{q}}^A
\right|^2 \right\}\,.
\label{pot}
\eeqn
Here  $f^{abc}$ denote the structure constants of the SU$(N)$ group,
$m_A$ is the mass term for the $A$-th flavor,
 and 
the sum over the repeated flavor indices $A$ is implied.
Above
we  introduced the FI $D$-term for the U(1) gauge factor with the FI 
parameter $\xi$. 

Now let us discuss  the vacuum structure of  this theory.
The  vacua of the theory (\ref{model}) are determined by the zeros of 
the potential (\ref{pot}). At generic values of the quark masses we have 
$$C_{N_f}^{N}= N_f!/N!\tN!$$ isolated $r$-vacua where $r=N$ quarks (out of $N_f$) develop
vacuum expectation values 
(VEVs).

Consider, say, the (1,2,...,$N$) vacuum in which the first $N$ flavors develop VEVs.
We can exploit gauge rotations to
make  all squark VEVs real. Then
in the problem at hand they take the form
\beqn
\langle q^{kA}\rangle &=&\sqrt{
\xi}\,
\left(
\begin{array}{cccccc}
1 & \ldots & 0 & 0 & \ldots & 0\\
\ldots & \ldots & \ldots  & \ldots & \ldots & \ldots\\
0 & \ldots & 1 & 0 & \ldots & 0\\
\end{array}
\right),
\qquad \langle \bar{\tilde{q}}^{kA}\rangle =0,
\nonumber\\[4mm]
k&=&1,..., N\,,\qquad A=1,...,N_f\, ,
\label{qvev}
\eeqn
where we write down the quark fields as  matrices in color and flavor indices.
This particular form of the squark condensates is dictated by first two lines
 in Eq.~(\ref{pot}). Note that the squark fields stabilize
at nonvanishing values exclusively due to the U(1) factor 
represented by the  term in the second  line.

The FI term $\xi$ singles  $r=N$ vacua out of  all set of $r$-vacua which
are present in the theory if quadratic in the adjoint field superpotential deformation $\mu{\mathcal A}^2$ is
added. In the vacuum under consideration the
adjoint fields also develop  
VEVs, namely,
\beq
\left\langle \left(\frac12\, a + T^a\, a^a\right)\right\rangle = - \frac1{\sqrt{2}}
 \left(
\begin{array}{ccc}
m_1 & \ldots & 0 \\
\ldots & \ldots & \ldots\\
0 & \ldots & m_N\\
\end{array}
\right),
\label{avev}
\eeq
For generic values of the quark masses, the  SU$(N)$ subgroup of the gauge 
group is
broken down to U(1)$^{N-1}$. However, in the special limit
\beq
m_1=m_2=...=m_{N_f},
\label{equalmasses}
\eeq
the  SU$(N)\times$U(1) gauge group remains  unbroken by the adjoint field.
In this limit the theory acquires a global flavor SU$(N_f)$ symmetry.

While the adjoint VEVs do not break the SU$(N)\times$U(1) gauge group in the limit
(\ref{equalmasses}), the quark condensate (\ref{qvev}) results in  the spontaneous
breaking of both gauge and flavor symmetries.
A diagonal global SU$(N)$ combining the gauge SU$(N)$ and an
SU$(N)$ subgroup of the flavor SU$(N_f)$
group survives, however. Below we will refer to this diagonal
global symmetry as to $ {\rm SU}(N)_{C+F}$.

More exactly, the pattern of breaking of the
color and flavor symmetry 
is as follows: 
\beq
{\rm U}(N)_{\rm gauge}\times {\rm SU}(N_f)_{\rm flavor}\to  {\rm SU}(N)_{C+F}\times  {\rm SU}(\tilde{N})_F\times {\rm U}(1)\,,
\label{c+f}
\eeq
where $\tilde{N}$ is defined in (\ref{tilen}).
Here SU$(N)_{C+F}$ is a global unbroken color-flavor rotation, which involves
first $N$ flavors, while the SU$(\tN )_F$ factor stands for the flavor rotation of the 
$\tN$ quarks.
The phenomenon of color-flavor locking takes place in the vacuum, albeit in a slightly different way 
than in the case $N_f = N$ (or $\tilde N =0$).
The presence of the global SU$(N)_{C+F}$ group is instrumental for
formation of the non-Abelian strings (see below).
 For unequal quark masses the  global symmetry  (\ref{c+f}) is broken down to 
U(1)$^{N_f-1}$.

Now let us discuss the mass spectrum in the theory (\ref{model}). Since
both U(1) and SU($N$) gauge groups are broken by squark condensation, all
gauge bosons become massive. From (\ref{model}) we get for the U(1)
gauge boson mass
\beq
m_{\gamma} =g_1\,\sqrt{\frac{N}{2}\,\xi\,}\, .
\label{phmass}
\eeq
At the same time, $(N^2-1)$ gauge bosons of the SU($N$) group acquire one and the same mass
\beq
m_{W}=g_2\sqrt{\xi}.
\label{wmass}
\eeq
It is not difficult to see from (\ref{pot}) that the adjoint fields $a$ and
$a^a$ as well as $N^2$ components of the quark matrix $q$ acquire
the same masses as the corresponding gauge bosons. 
Altogether we have one long \ntwo massive vector
multiplet (eight bosonic + eight fermionic states) with the mass
(\ref{phmass}) and $(N^2-1)$ long \ntwo massive vector multiplets with the mass (\ref{wmass}).
If the extra $\tN$ quark masses are different from 
those of the first $N$ masses (i.e. $m_{1,...,N}$), the extra quark
flavors acquire masses determined by the mass differences
$\Delta m_{PK}=m_P-m_K$, where $P=1, ..., N$ numerates the quark flavors which develop
VEVs in the $(1, ..., N)$ vacuum, while $K=N+1, ..., N_f$ numerates extra quark flavors. 
The extra flavors become massless in the
limit (\ref{equalmasses}), which we will consider momentarily.

Note that all states come in representations of the unbroken global
 group (\ref{c+f}), namely, the singlet and adjoint representations
of SU$(N)_{C+F}$
\beq
(1,\, 1), \quad (N^2-1,\, 1),
\label{onep}
\eeq
 and bifundamentals
\beq
 \quad (\bar{N},\, \tN), \quad
(N,\, \bar{\tN})\,,
\label{twop}
\eeq
 where we mark representation with respect to two 
non-Abelian factors in (\ref{c+f}).

\vspace{1mm}

If all quark mass terms are equal,  then all  $C_{N_f}^{N}$ isolated
vacua we had in the case of  unequal mass terms 
coalesce;  a Higgs branch develops from the common root whose location
on the Coulomb branch is given by Eq.~(\ref{avev}) with $\Delta m_{AB}=0$. The 
dimension of this branch is \cite{APS,MY}
\beq
{\rm dim}\, {\cal H}\,\Big|_{\xi\gg\Lambda}= \, 4 N N_f - 2 N^2 -N^2 -N^2 =4\tN N,
\label{dimH}
\eeq
where we take into account the fact that we have $4N N_f$ quark real degrees of freedom
and subtracted $2N^2 $  conditions due to $F$ terms, $N^2 $ conditions due to $D$ terms and, finally, 
$N^2 $ gauge phases eaten by the Higgs mechanism, see (\ref{pot}).

The Higgs branch is non-compact and is known to have a hyper-K\"ahler geometry \cite{SW2,APS}.
At a generic point on the Higgs branch BPS-saturated string solutions do not exist \cite{RuPe};
strings become non-BPS if we move along non-compact directions \cite{EY}.
However, the Higgs branch
has a compact base manifold defined by the condition
\beq
\tilde{q}_{Ak}=0\,,\qquad A=1, ...,\,N_f\,.
\label{tildeq}
\eeq
The dimension of this manifold 
is $2 N\tN$, twice less than the overall dimension of the Higgs branch.  
The BPS-saturated string solutions exist on the base manifold of the Higgs
branch. As a result, the vacua  belonging to the base manifold are our prime focus.

\vspace{1mm}

The base of the Higgs branch can be generated by flavor rotations of the 
(1,...,$N$) vacuum (\ref{qvev}). The flavor rotations generate the  manifold
\beq
\frac{{\rm SU}(N_f)}{{\rm SU}(N)_{C+F} \times {\rm SU}(\tN)\times U(1)},
\label{breakpattern}
\eeq
see Eq.~(\ref{c+f}). We see that the number of broken generators of the global group is
$2N\tN$. It coincides with the dimension of the base of the Higgs branch.

Since $N$ different flavors develop VEVs on the Higgs branch it is a baryonic 
Higgs branch. It is a generalization of the baryonic Higgs branch  \cite{APS}
to the case of the U($N$) gauge group and nonvanishing masses. Note, however, that in 
the U($N$) gauge theory the  baryonic charge is gauged, in contradistinction with \cite{APS}.

\vspace{1mm}

Now let us have a closer look at quantum effects in the theory
(\ref{model}). The SU$(N)$ sector is
asymptotically free. 
The semiclassical analysis outlined above is valid if the FI parameter $\xi$ 
is large,
\beq
\xi\gg \Lambda\, ,
\label{weakcoupling}
\eeq
where $\Lambda$ is the dynamical scale of the SU($N$) gauge theory.
This condition ensures weak coupling in the SU$(N)$ sector because
the SU$(N)$ gauge coupling does not run below the scale of the quark VEVs
which is determined by $\xi$. More explicitly,
\beq
\frac{8\pi^2}{g^2_2 (\xi)} =
(N-\tN )\ln{\frac{g_2\sqrt{\xi}}{\Lambda}}\gg 1 \,.       \rule{8mm}{0mm}
\label{4coupling}
\eeq

\vspace{2mm}

Below we will see   that if we pass to small  $\xi$ following the line $\Delta m_{A,B}=0$, 
into the 
strong coupling domain III,  where  the condition (\ref{weakcoupling}) is not met,
the theory undergoes a crossover.
In the case $N_f=N$ studied in \cite{SYcross} this is a transition  into the 
Seiberg--Witten Abelian 
regime. In this regime  no non-Abelian strings develop. We will show below
that if $N_f>N$ the theory at small $\xi$ below the transition point at $\xi\sim\Lambda^2$
is still non-Abelian, with the dual gauge group U($\tN$). It supports non-Abelian semilocal strings  
for which the role of orientation and size moduli is interchanged.

To conclude this section we briefly recall the theory ({\ref{model})
at nonvanishing quark mass differences $m_A-m_B\neq 0$, see \cite{SYmon,SYrev}.
At $m_A-m_B\neq 0$ the global group (\ref{c+f}) is 
explicitly broken down to U(1)$^{N_f-1}$. The adjoint multiplet is split. The diagonal entries 
(photons and their \ntwo quark superpartners) have
masses given in (\ref{wmass}), while the off-diagonal states ($W$ bosons and the
off-diagonal entries of the squark matrix  $q^{kA}$ with $A\neq k$) acquire additional contributions 
to their masses  proportional to $\Delta m_{AB}$. In particular,
$\tN$ ``extra" quark flavors become massive, and the Higgs branch is lifted. 
As we make the mass differences larger, the
$W$ bosons become exceedingly  heavier,  decouple from
the low-energy spectrum,  and we are left with $N$ photon states and
 $N$ diagonal elements of the quark matrix with $A= k$. The low-energy spectrum
becomes Abelian.

\subsection {Non-Abelian strings at large \boldmath{$\xi$}}
\label{strings}

Now we will briefly review non-Abelian strings  in the theory (\ref{model}),
see  \cite{SYrev} for details. Non-Abelian strings in \ntwo QCD with $$N_f=N$$
where first found and studied in \cite{HT1,ABEKY,SYmon,HT2}.
The Abelian $Z_N$-string solutions break the  SU$(N)_{C+F}$ global group. Therefore,
strings have orientational zero modes, associated with rotations of their color 
flux inside the non-Abelian SU($N$). This makes these strings non-Abelian.
The global group is broken on the $Z_N$ string solution down to 
${\rm SU}(N-1)\times {\rm U}(1)$.
As a result,
the moduli space of the non-Abelian string is described by the coset
\beq
\frac{{\rm SU}(N)}{{\rm SU}(N-1)\times {\rm U}(1)}\sim {\rm CP}(N-1)\,.
\label{modulispace}
\eeq
The CP$(N-1)$ space can be parametrized by
 a complex vector $n^P$
 in the fundamental representation of SU($N$) subject to the constraint
 $$n^*_P n^P ={\rm const}\,,$$
where $P=1, ..., N$.
As we will show below, one U(1) phase will be gauged away in the low-energy
sigma model. This gives the correct number of degrees of freedom,
namely, $2(N-1)$.

Making the moduli vector $n^P$ a slowly varying function of the string world
sheet coordinates $x_{\alpha}$ ($\alpha=0,3$), we can derive an
effective low-energy theory on the string world sheet 
\cite{ABEKY,SYmon,GSY05}. On topological grounds
(see (\ref{modulispace})) it is clear that we will get the
two-dimensional CP$(N-1)$ model.
The \ntwot supersymmetric CP$(N-1)$ model can be understood as
a strong-coupling limit  of a U(1) gauge theory \cite{W93}. The  bosonic part
of the  action of this model has the form
\beqn
S_{{\rm CP}(N-1)}
& =&
\int d^2 x \left\{
\left|\nabla_{\alpha} n^{P}\right|^2 +\frac1{4e^2}F^2_{\alpha\beta} + \frac1{e^2}
|\pt_\alpha\sigma|^2
\right.
\nonumber\\[3mm]
 &+&    2\,\left|\sigma+\frac{m_P}{\sqrt{2}}\right|^2 |n^{P}|^2 + 
\frac{e^2}{2} \left(|n^{P}|^2 -2\beta\right)^2
\Big\}\,,
\label{cpg}
\eeqn
where $\nabla_{\alpha}= \partial_{\alpha} - i A_{\alpha} $ while $\sigma$ is a complex scalar
field, and summation over $P$ is implied. The condition 
\beq
 n^*_P n^P =2\beta\,,
\label{unitvec}
\eeq 
is implemented in the limit $e^2\to\infty$. Moreover, in this limit
the gauge field $A_{\alpha}$  and its \ntwo bosonic superpartner $\sigma$ become
auxiliary and can be eliminated by virtue of the equations of motion,
\beq
A_{\alpha} =-\frac{i}{4\beta}\, n^*_P \stackrel{\leftrightarrow}
{\partial_{\alpha}} n^P \,,\qquad \sigma=0\,.
\label{aandsigma}
\eeq
The two-dimensional coupling constant $\beta$  is determined by the
four-dimensional non-Abelian coupling via the relation
\beq
\beta= \frac{2\pi}{g_2^2}\,.
\label{betag}
\eeq

\vspace{2mm}

In the limit of equal quark masses the global SU($N)_{C+F}$ symmetry is unbroken, and 
strings become non-Abelian. This is a strong coupling quantum regime
in the CP$(N-1)$ model (\ref{cpg}). The vector $n^P$ is smeared all over the entire CP$(N-1)$
space due to quantum fluctuations and its average value vanishes \cite{W79}.
 The world-sheet theory develop a mass gap.

At small nonvanishing $\left|m_P-m_{P'}\right|$ the global SU($N)_{C+F}$ symmetry is 
explicitly broken down to $U(1)^{(N-1)}$.
A shallow potential is generated on the  CP$(N-1)$ modular space as is seen from
(\ref{cpg}). As we increase $\left|m_P-m_{P'}\right|$ the strings become exceedingly more Abelian
and eventually evolve into Abelian $Z_N$ strings, which correspond to
$N$ classical vacua of the world sheet model (\ref{cpg})
\beq
n^P=\sqrt{2\beta}\;\delta^{PP_0},\qquad \sigma=-\,\frac{m_{P_0}}{\sqrt{2}},
\label{cphiggsvac}
\eeq
where $P_0$ can take any of $N$ values, $P_0=1 , ..., N$, see the review \cite{SYrev}. Note, that we should keep
mass differences $(m_P-m_{P'})$ small as compared to the inverse string thickness,
\beq
\left|m_P-m_{P'}\right|\ll g\sqrt{\xi}\,,
\label{smallmP}
\eeq
where we assume that $g_1\sim g_2\sim g$. 

The CP$(N-1)$ model is 
an effective low -energy description of the internal string dynamics, and
the bulk mass scale $g\sqrt{\xi}$ plays the role of an ultraviolet (UV) cut-off in (\ref{cpg}).
The constraint (\ref{smallmP}) ensures that typical energies in the world-sheet theory
are much lower then this UV cut-off.

Let us ask ourselves what happens  if we add ``extra" quark flavors with degenerate mass?
Then the strings emerging in the theory with $N_f>N$ become semilocal.
In particular, the string solutions on the Higgs branches (typical
for multiflavor theories) usually are not fixed-radius strings, but, rather,
semilocal strings, see  the review paper \cite{AchVas} for a comprehensive survey of Abelian semilocal strings.

Let us start our discussion with such Abelian semilocal strings.
The semilocal string interpolates between the Abrikosov--Nielsen--Olesen string 
\cite{ANO} and two-dimensional  sigma-model instanton
lifted to four dimensions (this is referred to as
lump). The relevance of instantons can be understood as follows. We can go to 
low energies (below the photon mass (\ref{phmass})) and then integrate out massive states.
In this limit the theory reduces to a sigma model on the Higgs branch.
If we stay at the base of the Higgs branch imposing condition (\ref{tildeq}), and
this base has an $S_2$ cycle, the theory has lumps.
Much in the same way as the instanton/lump, the semilocal string  possesses
 additional zero modes associated with complexified string's transverse size $\rho$.
At  $\rho\to 0$ we have the ANO string while
at $\rho \to\infty$ it becomes a pure lump.
At $\rho \neq 0$ the profile functions of the semilocal
string fall-off at infinity as inverse powers
of the distance to the string axis,
instead of the exponential fall-off characteristic to the ANO strings at $\rho=0$.
This leads to a dramatic physical effect ---  
semilocal strings, in contradistinction to the ANO ones,
do not support linear confinement \cite{EY,SYsem}.

Non-Abelian semilocal strings in \ntwo QCD with $N_f>N$ were studied in
\cite{HT1,HT2,SYsem,Jsem}. These strings have both types of moduli: orientational and size moduli.
The orientational 
zero modes of the semilocal non-Abelian string are parametrized by the complex vector $n^P$, $P=1,...,N$,
 while its $\tN$ size moduli are parametrized by the  complex vector
$\rho^K$, $K=N+1,...,N_f$. The effective two-dimensional theory
which describes the internal dynamics of the non-Abelian semilocal string is
an \ntwot ``toric" sigma model which includes both types of fields. Its bosonic action
in the gauge formulation (which assumes taking the limit $e^2\to\infty$) has the form
\beqn
&&S = \int d^2 x \left\{
 \left|\nabla_{\alpha} n^{P}\right|^2 
 +\left|\tilde{\nabla}_{\alpha} \rho^K\right|^2
 +\frac1{4e^2}F^2_{\alpha\beta} + \frac1{e^2}\,
\left|\pt_{\alpha}\sigma\right|^2
\right.
\nonumber\\[3mm]
&+&\left.
2\left|\sigma+\frac{m_P}{\sqrt{2}}\right|^2 \left|n^{P}\right|^2 
+ 2\left|\sigma+\frac{m_{K}}{\sqrt{2}}\right|^2\left|\rho^K\right|^2
+ \frac{e^2}{2} \left(|n^{P}|^2-|\rho^K|^2 -2\beta\right)^2
\right\},
\nonumber\\[4mm]
&& 
P=1,...,N\,,\qquad K=N+1,...,N_f\,,\qquad \tilde{\nabla}_k=\pt_k+iA_k\,.
\label{wcp}
\eeqn
The fields $n^{P}$ and $\rho^K$ have
charges  +1 and $-1$ with respect to the U(1) gauge field,
hence, the difference in the covariant derivatives, $\nabla_{\alpha}$ and $\tilde{\nabla}_{\alpha}$, respectively.

If only charge $+1$ fields were present, in the limit 
$e^2\to\infty$ we would get a conventional twisted-mass deformed
CP($N-1$) model.
The presence of the charge $-1$ fields $\rho^K$ converts 
the target space of the CP$(N-1)$ sigma model into a weighed CP$(N_f-1)$ space. 
Like in CP$(N-1)$ model (\ref{cpg}), small mass differences
$\left| m_A-m_B\right|$ lift orientational and size zero modes generating a shallow potential
on the modular space.

The world-sheet theory (\ref{wcp}) was argued to emerge as an effective low-energy
theory on the world sheet of the semilocal non-Abelian string in \cite{HT1,HT2}.
The arguments were based on a $D$-brane construction. Later this result was
confirmed by direct derivations from the bulk theory in \cite{SYsem,Jsem}.
These derivations have a subtle point, though. Both orientational and size moduli
have a logarithmically divergent in the infrared (IR) norm in the limit $\Delta m_{AB}=0$.
This divergence is cut off by small mass
differences $\left| m_P-m_K\right|\neq 0$ (here $P=1,...,N$ and $K=N+1,...,N_f$). 
What counts is the difference between the masses of $N$ quarks 
which develop 
VEVs in the bulk vacuum and the masses of ``extra" $\tN$ quarks. 
With this cut-off,  The large
logarithmic factor can be absorbed in the field definition \cite{SYsem}.
The theory (\ref{wcp}) emerges in a logarithmic approximation in  which this logarithmic factor is 
large. This ensures that  $|\rho|\ll g\sqrt{\xi}/\left| m_P-m_K\right|$. 

The  two-dimensional coupling constant $\beta$ is related to the four-dimensional
one via (\ref{betag}).
This relation  is obtained  at the classical level 
\cite{ABEKY,SYmon}.
 In quantum theory
both couplings run. In particular, the  model (\ref{wcp}) is asymptotically free
\cite{W93} and develops its own scale $\Lambda_{\sigma}$. 
The ultraviolet cut-off in the sigma model on the string world sheet
is determined by  $g\sqrt{\xi}$. 
Equation~(\ref{betag}) relating the two- and four-dimensional couplings
is valid at this scale.
At this scale the four-dimensional coupling is given by (\ref{4coupling})
while the two-dimensional one
\beq
4\pi\beta(\xi) =
\left(N-\tN \right)\ln\,{\frac{g\sqrt{\xi}}{\Lambda}}\gg 1 \,.
\label{2coupling}
\eeq
Then Eq.~(\ref{betag}) implies
\beq
\Lambda_{\sigma} =  \Lambda\, .
\label{lambdasig}
\eeq

Note that in the bulk theory {\em per se} the coupling constant is frozen at
$g_2\sqrt{\xi}$, because of the VEVs of
the squark fields.
The logarithmic evolution of the coupling constant in the string world-sheet theory takes over.
Moreover, the dynamical scales of the bulk and world-sheet
theories turn out to be the same, much in the same way as in the $N_f=N$ theory \cite{SYmon}.

\section{The  bulk duality}
\label{bulkdual}
\setcounter{equation}{0}

Our task in this section is to analyze the transition  from the domain I to 
III (see Fig.~\ref{figphasediag}). This will be done in two steps.
First
we will take  the quark mass differences to be large, passing to the domain II. In this domain the theory stays at
weak coupling, and we can safely diminish the value of the FI parameter $\xi$. Next,
we will use the exact Seiberg--Witten solution of the theory on the Coulomb branch \cite{SW1,SW2}
(i.e. at   $\xi=0$) to perform the passage from the domain II to  III. 

\subsection{The dual gauge group}
\label{secdualgaugegroup}

To begin with, let us identify the $r=N$ quark vacuum of the form $(1, ... , N)$
which was described  above semiclassically. To this end we will use the exact
Seiberg--Witten solution \cite{SW1,SW2}, more exactly, the SU$(N)$ generalizations
of the Seiberg--Witten solution \cite{ArFa,KLTY,ArPlSh,HaOz}.

Instead of considering generic quark masses we will make a representative (and convenient) choice.
Then we will show that
the low-energy effective theory at small $\xi$ and small quark mass differences (in the domain III)
has the dual non-Abelian gauge group U($\tN$). 

Our special choice of the quark masses
ensures that this theory is not asymptotically free -- in fact, it is IR free -- and stays  at weak coupling
at small $\xi$, cf. Ref.~\cite{APS}.  The set of masses we will deal with in this section is as follows:
the masses of the extra $\tN$ quark fields are to be set equal to
the masses of the first $\tN$ quarks from those $N$ squarks which develop VEVs in the
(1,...,$N$) vacuum. Namely, we set
\beq
m_1=m_{N+1}\,,\quad  m_2=m_{N+2}\,, ... ,\,  m_{\tN} =m_{N+\tN}\,.
\label{masschoice}
\eeq
Later on we will be able to relax these conditions.
The Seiberg--Witten curve in the theory under consideration has the form \cite{APS}
\beq
y^2= \prod_{k=1}^{N} (x-\phi_k)^2 -
4\left(\frac{\Lambda}{\sqrt{2}}\right)^{N-\tN}\, \,\,\prod_{A=1}^{N_f} \left(x+\frac{m_A}{\sqrt{2}}\right),
\label{curve}
\eeq
where $\phi_k$ are gauge invariant parameters on the Coulomb branch. Semiclassically,
at large masses
\beq
{\rm diag}\left(\frac12\, a + T^a\, a^a\right) \approx 
\left[\phi_1,...,\phi_N\right].
\eeq
Therefore, in  the ($1, ... ,\, N$) quark vacuum we have
\beq
\phi_P \approx -\frac{m_P}{\sqrt{2}},\qquad P=1, ... ,\, N\,,
\label{classphi}
\eeq
in the large $m_A$ limit, see (\ref{avev}). 

To identify this vacuum in terms of the curve (\ref{curve}) it is necessary to find
such values of $\phi_P$ which ensure that the curve has $N$ double roots and
$\phi_P$'s are determined by the quark masses in the semiclassical limit, see (\ref{classphi}).
For the mass choice (\ref{masschoice}) the solution can be  easily obtained. Indeed, let us write
the curve in the form
\beqn
y^2
&=&
 \prod_{P=1}^{\tN} \left(x+\frac{m_P}{\sqrt{2}}\right)^2 
\nonumber\\[3mm]
&\times&
\left\{\prod_{k=\tN+1}^{N} (x-\phi_k)^2-
4\left(\frac{\Lambda}{\sqrt{2}}\right)^{N-\tN}\, \prod_{P=\tN+1}^{N} \left(x+\frac{m_P}{\sqrt{2}}\right)\right\},
\label{redcurve}
\eeqn
where the first $\tN$ $\phi$'s are given by
\beq
\phi_P = -\frac{m_P}{\sqrt{2}},\qquad P=1,...,\tN.
\label{firstphi}
\eeq
This curve has $\tN$ double roots located at
\beq
x_P =-\frac{m_P}{\sqrt{2}},\qquad P=1, ... ,\,\tN.
\label{firstroots}
\eeq

Now to find other double roots and $\phi$'s we have to investigate the reduced curve
in the curly brackets in (\ref{redcurve}). It corresponds to the U($N-\tN$) gauge theory with
$(N-\tN)$ flavors. This theory completely Abelianizes below the crossover transition 
(at small $\xi$) \cite{SYcross}. In other words, the corresponding $\phi$'s get  shifts
from their classical values (\ref{classphi}) proportional to $\Lambda$. To see this 
explicitly let us consider the  simplest special case with all ``extra" $(N-\tN)$
masses are equal,
\beq
m_P=m, \qquad P=(\tN+1) , ... ,\, N
\label{extraeqmass}
\eeq
and $(N-\tN)=2^p$ (the latter condition is imposed for simplicity). 
Then the curve (\ref{redcurve}) reduces to a perfect square
\beq
y^2= \prod_{P=1}^{\tN} \left(x+\frac{m_P}{\sqrt{2}}\right)^2 \left\{ 
\left(x+\frac{m}{\sqrt{2}}\right)^{N-\tN}-\left(\frac{\Lambda}{\sqrt{2}}\right)^{N-\tN}\right\}^2
\label{perfsquare}
\eeq
provided that
\beq
\phi_k=\frac1{\sqrt{2}}\,\left[-m_1,...,-m_{\tN},-m+
\Lambda\,e^{\frac{\pi i}{N-\tN}},...,-m+
\Lambda\,e^{\frac{2\pi i}{N-\tN}(N-\tN-\frac12)}\right].
\label{phi}
\eeq
The first $\tN$ double roots are  given in Eq.~(\ref{firstroots})
while the remaining $N-\tN $ double roots are
\beq
x_P =\frac1{\sqrt{2}}\,\left[... ,\, -m+\Lambda,
... ,\, -m+\Lambda\,e^{\frac{2\pi i}{N-\tN}(N-\tN-1)}\right]
\label{roots}
\eeq
The main feature of this solution is the absence of $\sim\Lambda$ corrections to the first
$\tN$ $\phi$'s in (\ref{phi}). This means that in the equal mass limit these $\tN$ $\phi$'s
become equal. This is a signal of restoration of the non-Abelian U($\tN$) gauge group
at the root of the Higgs branch (i.e. at $\xi=0$). Namely, the gauge group at the root of
the Higgs branch in the equal mass limit becomes
\beq
{\rm U}(\tN)\times {\rm U}(1)^{N-\tN}.
\eeq
This is in a perfect agreement with the
results obtained in \cite{APS} where a dual non-Abelian gauge group  was identified
at the root of a baryonic Higgs branch in the SU($N$) gauge theory with massless quarks.
The novel element of our analysis in this section is that
we started with the $r=N$ non-Abelian vacuum at large $\xi$ and demonstrated that,
as we reduce $\xi$,
the theory in this vacuum undergoes crossover to another non-Abelian regime
with the dual low-energy gauge group (\ref{dualgaugegroup}).
As was already mentioned, the physical 
reason for the emergence of the non-Abelian  gauge group is that the low-energy
effective theory  with the dual gauge group (\ref{dualgaugegroup}) is not asymptotically free 
in the equal mass 
limit and stays at weak coupling. Therefore, the classical analysis showing that the non-Abelian
gauge group is restored at the root of the Higgs branch remains intact in
quantum theory.

\subsection{Monodromies}
\label{monodro}

In this section we will study how quantum numbers of the massless quarks $q^{11}, ... ,\, q^{NN}$
in the $(1, ... ,\, N)$ vacuum
change as we reduce $\Delta m_{AB})$ to pass from the domain II in 
III (the along Coulomb branch at $\xi=0$), where the theory is at strong coupling. 
To simplify our subsequent discussion we will consider 
a particular case: the  theory with $$N=3\,, \qquad N_f=5\,$$ 
so that the dual group has the smallest nontrivial rank $\tN=2$. We 
will consider the $(1, 2, 3)$ vacuum. In addition, we will
stick to  a 
special choice of the quark masses (\ref{masschoice}), which in the case at hand implies
\beq
m_1= m_4\,, \qquad m_2= m_5\,.
\label{masschu3}
\eeq
The mass parameter $m_3$ remains unspecified for the time being.

The quark quantum numbers change  due to monodromies 
with respect to $\Delta m_{PP'}\,$. The 
complex planes of $\Delta m_{PP'}$ have cuts, and when we cross these cuts, $a$ and $a_D$
fields acquire monodromies; the quantum numbers of the corresponding states change accordingly.
Monodromies with respect to the quark masses were studied in \cite{BF} in the theory
with the SU(2) gauge group through a monodromy matrix approach.

Here we will investigate the monodromies in the U(3) theory with five quark flavors
using the  approach of Ref.~\cite{SYcross} which is  similar to that of Ref.~\cite{CKM}. 
In the case $N_f=2N-1$, the Seiberg--Witten curve, instead of (\ref{curve}),
is given by \cite{APS}
\beq
y^2= \prod_{k=1}^{3} (x-\phi_k)^2 -
4\left(\frac{\Lambda}{\sqrt{2}}\right) \,\,\prod_{A=1}^{5} \left(x+\frac{\tilde{m}_A}{\sqrt{2}}\right),
\label{curvep}
\eeq
where, according to \cite{APS}, ``shifted" masses
\beq
\m_A \equiv m_A +\frac{\Lambda}{3}
\label{shift}
\eeq
replace $m_A$. Substituting (\ref{masschu3}) 
and 
\beq
\phi_1 = -\frac{\m_1}{\sqrt{2}},\qquad \phi_2 = -\frac{\m_2}{\sqrt{2}}\,,
\label{phi12}
\eeq
we arrive at
\beq
y^2= \left(x+\frac{\m_1}{\sqrt{2}}\right)^2 \left(x+\frac{\m_2}{\sqrt{2}}\right)^2 
\left[(x-\phi_3)^2  - 
 4\,\frac{\Lambda}{\sqrt{2}}\, \left(x+\frac{\m_3}{\sqrt{2}}\right)\right].
\label{U3curve}
\eeq
The first two double roots of this curve are obviously located at
\beq
e_1 = e_2=-\frac{\m_1}{\sqrt{2}},\qquad e_3 = e_4=-\frac{\m_2}{\sqrt{2}},
\label{roots1-4}
\eeq
cf. Eq.~(\ref{firstroots}).
The remaining two roots coincide provided we set
\beq
\phi_3=-\frac{1}{\sqrt{2}}\,(\m_3+\Lambda)\, .
\label{phi3}
\eeq
If we do so, the last two coinciding roots are  
\beq
e_5 = e_6=-\frac{1}{\sqrt{2}}\,(\m_3-\Lambda),
\label{roots56}
\eeq
cf. Eqs.~(\ref{phi}) and (\ref{roots}).

If two roots of the Seiberg--Witten curve coincide, the contour
which encircles  these roots shrinks and produces a regular potential.
We start from the quasiclassical regime at $\Delta m_{PP'}\gg \Lambda$. We have three double roots $e_1=e_2$,
 $e_3=e_4$  and $e_5=e_6$ in the $r=3$ vacuum. Thus, three contours $\alpha_1$, $\alpha_2$ 
and $\alpha_3$
shrink (see Fig.~\ref{figcontoursU3}), and the associated potentials $a$, $a_3$ and $a_8$ are regular. This is 
 related to masslessness of three quarks $q^{11}$, $q^{22}$ and $q^{33}$ (at $\xi=0$), 
\beqn
&&\frac12 \,a+ \frac12 \,a_3+ \frac1{2\sqrt{3}}\, a_8 +\frac{m_1}{\sqrt{2}}=0\,,\nonumber\\[2mm]
&&\frac12 \,a- \frac12 \,a_3+ \frac1{2\sqrt{3}} \,a_8 +\frac{m_2}{\sqrt{2}}=0\,,\nonumber\\[2mm]
&&\frac12\, a -\frac1{\sqrt{3}}\, a_8 +\frac{m_3}{\sqrt{2}}=0 \,.
\label{masslessquarks}
\eeqn
Here we exploit the fact that the charges 
of these three quarks are as follows:
\beqn
&&\left(n_e,n_m;\,n_e^3,n_m^3;\,n_e^8,n_m^8\right)=
\left(\frac12,0;\,\frac12,0;\,\frac1{2\sqrt{3}},0\right), \nonumber\\[2mm]
&&\left(n_e,n_m;\,n_e^3,n_m^3;\,n_e^8,n_m^8\right)=
\left(\frac12,0;\,-\frac12,0;\,\frac1{2\sqrt{3}},0\right), \nonumber \\[2mm]
&&\left(n_e,n_m;\,n_e^3,n_m^3;\,n_e^8,n_m^8\right)=
\left(\frac12,0;\,0,0;\,-\frac1{\sqrt{3}},0\right),
\label{quarkcharges}
\eeqn
respectively, where
$n_e$ and $n_m$ denote electric and 
magnetic charges of a given state with respect to the U(1) gauge group, while $n_e^3$,
$n_m^3$ and $n_e^8$, $n_e^8$ stand for the electric and magnetic charges  with respect to the Cartan
generators of the SU(3) gauge group (broken down to U(1)$\times$U(1) by $\Delta m_{PP'}$).

\begin{figure}
\epsfxsize=9cm
\centerline{\epsfbox{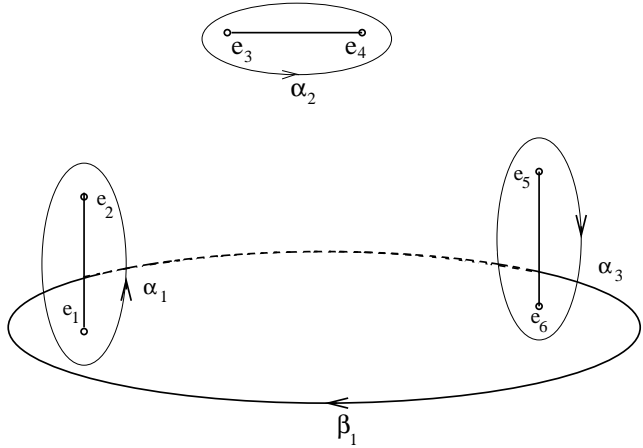}}
\caption{\small Basis of $\alpha$ and $\beta$ contours in U(3) gauge theory.
Two roots $e_3$ and $e_4$ are far away near AD point (\ref{AD1}).}
\label{figcontoursU3}
\end{figure}

In the monopole singularity certain other roots coincide. Say, $e_1=e_6$,
see Fig.~\ref{figcontoursU3}. Thus,
the $\beta_1$ contour
shrinks producing a regular $\frac12\, a^D_3+\frac{\sqrt{3}}{2}\,a^D_8$
potential. This is due to masslessness of the monopole (one  of three  SU(3) monopoles) with 
the charges\,\footnote{The charges of three elementary SU(3) monopoles are determined by 
the roots of the SU(3) Cartan subalgebra.}
\beq
\left(n_e,n_m;\,n_e^3,n_m^3;\,n_e^8,n_m^8\right)=
\left(0,0;\,0,\frac12;\,0,\frac{\sqrt{3}}{2}\right).
\label{1moncharge}
\eeq

If we decrease $|\Delta m_{PP'}|$  crossing  cuts in the $\Delta m_{PP'}$ planes,
 the root pairing
 in the given vacuum may change. This would mean that  other
combinations of $a$'s and $a^D$'s become regular implying a change of 
the quantum numbers
of the massless states in the  given vacuum. To see how it works in our $r=3$ vacuum
let us go to the Argyres--Douglas (AD) point\,\cite{AD,APSW}.
The AD point is a  particular value of the quark mass parameters where more mutually nonlocal states
become massless. In fact, we will study the collision of the
$r=3$ quark vacuum
with  monopole singularities. We approach the AD points from the domain II
at large $|\Delta m_{PP'}|$. We will show below that as we pass through the AD points
the root pairings   change in the $r=3$ vacuum implying a change of the 
quantum numbers of the massless states. Three massless quarks transform into three
massless dyons. 

From (\ref{roots1-4}) and (\ref{roots56}) we see that there are two AD points where our $(1,2,3)$
vacuum collides with the monopole singularities. The first one occurs at  
\beq
\Delta m_{31}=\Lambda, \qquad 
e_1=e_2=e_5=e_6=-\frac{\m_1}{\sqrt{2}}\,,
\label{AD1}
\eeq
where four roots coincide, while the second is at
\beq
\Delta m_{32}=\Lambda, \qquad e_3=e_4=e_5=e_6=-\frac{\m_2}{\sqrt{2}}\,,
\label{AD2}
\eeq
where other four roots coincide.

We assume that $m_1$ and  $m_2$ are real, $m_1>m_2$, and 
consider the first AD point (\ref{AD1}). At this point the $(1,2,3)$ vacuum with three massless quarks
(\ref{quarkcharges}) collide with the monopole singularity where the monopole (\ref{1moncharge})
is massless. We will demonstrate below that as we reduce $\Delta m_{31}$ 
along the real axis below the AD point (\ref{AD1}) the root
pairings change. The roots $e_3$ and $e_4$ are far away and, therefore, the charges of the $q^{22}$ quark
 do not change. We focus on the colliding roots $e_1$, $e_2$, $e_5$ and  $e_6$.

In order to see how the root pairings   in the $r=3$ vacuum change as we decrease 
$\Delta m_{31}$ and pass  through the AD point
(\ref{AD1}), we have to slightly split the roots by shifting $\phi_1$ 
from its $r=3$ solution (\ref{phi12}). Let us parametrize the shift as
\beq
\phi_1 = -\frac{\m_1}{\sqrt{2}} + \frac{\delta}{4\Lambda^2}\,,
\label{delta}
\eeq
where $\delta$ is a small deviation parameter of mass dimension three.
Since we will consider $x$ in the vicinity of $-\tilde{m}_1/\sqrt{2}$ we  introduce 
another small parameter $z$,
\beq
z=x+ \frac{\m_1}{\sqrt{2}} \,.
\label{xshift}
\eeq
Finally, we  define 
\beq
\varepsilon=\frac{\Delta m_{31}-\Lambda}{\sqrt{2}}\,.
\eeq
The parameter $\varepsilon$ is a small deviation from the AD point (\ref{AD1}).
Furthermore, we will expand (\ref{curvep}) in $\delta$, omitting terms
$O(\delta^2)$, $O(\delta z^2 )$ and $O(z \delta\varepsilon)$. The factor $(x+\frac{\m_2}{\sqrt{2}})^2$ 
can and will be approximated by $\Delta m_{12}^2/2$.
Then the curve (\ref{curvep}) takes the form
\beq
y^2\approx \frac{\Delta m_{12}^2}{2}\left[z^2(z+\varepsilon)^2-z\delta\right].
\label{AD1curve}
\eeq
 
 \vspace{2mm}
 
Above the AD point, at 
$\varepsilon>0$ and $\delta\ll\varepsilon$,
the roots of the curve (\ref{AD1curve}) are split as follows:
\beqn
&& z_1=0\,,\qquad z_2=\frac{\delta}{\varepsilon^2}\,,
 \nonumber\\[2mm]
&& z_5=-\varepsilon -\sqrt{-\frac{\delta}{\varepsilon}}\,,\qquad 
z_6=-\varepsilon +\sqrt{-\frac{\delta}{\varepsilon}}\,,
\eeqn
where $z_i$ are shifted $e_i$ (see (\ref{xshift})), i.e. $$z_i = e_i + (\tilde{m}_1/\sqrt{2})\,.$$ 
We take $-i\delta>0$ and 
study  the evolution of the roots of the curve
(\ref{AD1curve}) as a function of $\varepsilon$
numerically. The results are schematically presented in 
Fig.~\ref{figrootsU3}.
We see that the root pairings in the $r=3$ vacuum change. Namely,
at large $\Delta m_{31}$ we have (at $\delta=0$)
\beq
e_1=e_2, \qquad e_5=e_6, \qquad e_3=e_4
\label{IIroots}
\eeq
 which, as was explained above, corresponds to shrinking of the
$\alpha_1$, $\alpha_2$ and $\alpha_3$ contours and masslessness of three quarks
(\ref{quarkcharges}).
\begin{figure}
\epsfxsize=10cm
\centerline{\epsfbox{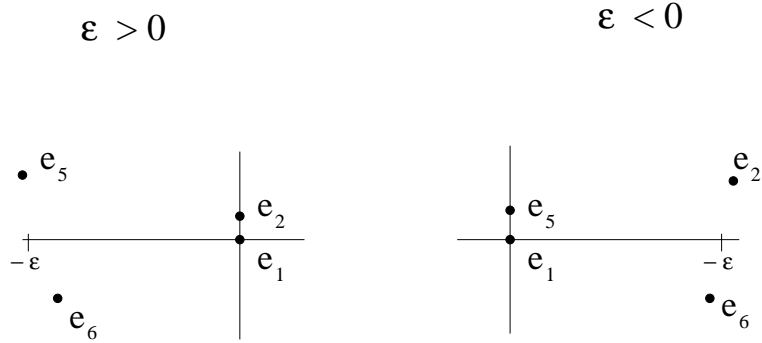}}
\caption{\small As we decrease $\Delta m_{31}$  and pass through the
AD point (\ref{AD1}), the pairing of roots $e_{1,2,5,6}$ in the $x$ plane changes.}
\label{figrootsU3}
\end{figure}
Below the AD point (\ref{AD1}), at small $\Delta m_{31}$, we have 
\beq
e_1=e_5\,, \qquad e_2=e_6\,,\qquad e_3=e_4\,,
\label{IIIroots}
\eeq
which corresponds to shrinking of the contours
\beq
\alpha_1+\beta_1\to 0,\qquad \alpha_3 -\beta_1\to 0\qquad \alpha_2\to0\,,
\label{shrcont}
\eeq
see Fig.~\ref{figcontoursU3}.
This means that the massless quarks  $q^{11}$ and $q^{33}$ in the $r=3$ vacuum transform
themselves into
massless  dyons $D^{11}$ and $D^{33'}$, with the quantum numbers
\beqn
&&D^{11}:\,\,\, \left(\frac12,0;\,\frac12,\frac12;\,\frac1{2\sqrt{3}},\frac{\sqrt{3}}{2}\right),
 \nonumber\\[3mm] 
 && D^{33'}:\,\,\,\,
\left(\frac12,0;\,0,-\frac12;\,-\frac1{\sqrt{3}},-\frac{\sqrt{3}}{2}\right),
\label{dyons1}
\eeqn
while the charges of the quark $q^{22}$ do not change.
We see that the quantum numbers of the massless quarks $q^{11}$ and $q^{33}$ in the 
$r=3$ vacuum, after the 
collision with the monopole singularity, get shifted, the shift being equal to
$\pm$(monopole magnetic charge). 

By the same token, we can analyze the second AD point (\ref{AD2}) where the $r=3$ vacuum
collides with another monopole singularity in which the monopole with the charges 
\beq
\left(0,0;\,0,-\frac12;\,0,\frac{\sqrt{3}}{2}\right)
\label{2moncharge}
\eeq
is massless. The corresponding results are as follows: 
now $D^{11}$ does not change its charges, while the
charges of the quark $q^{22}$ and dyon $D^{33'}$ get a shift by the $\pm$(charge of the monopole
(\ref{2moncharge})).
As a result, below the AD point (\ref{AD2}) the charges of the massless dyons are
\beqn
&& D^{11}:\,\,\, \left(\frac12,0;\,\frac12,\frac12;\,\frac1{2\sqrt{3}},\frac{\sqrt{3}}{2}\right),
\nonumber\\[2mm]
&& D^{22}:\,\,\, 
\left(\frac12,0;\,-\frac12,-\frac12;\,\frac1{2\sqrt{3}},\frac{\sqrt{3}}{2}\right),
\nonumber\\[2mm]
&& D_{33}:\,\,\, \,
\left(\frac12,0;\,0,0;\,-\frac1{\sqrt{3}},-\sqrt{3}\right).
\label{dyons}
\eeqn
The quark masslessness conditions  (\ref{masslessquarks}) at small $\Delta m_{PP'}$,
below two AD points, are replaced by dyon masslessness conditions, namely,
\beqn
&&\frac12 \,a+ \frac12 \,a_3+ \frac12 \,a^D_3+ \frac1{2\sqrt{3}}\, a_8
+\frac{\sqrt{3}}{2}\, a^D_8 +\frac{m_1}{\sqrt{2}}=0\,,
\nonumber\\[2mm]
&&\frac12 \,a- \frac12 \,a_3 - \frac12 \,a^D_3+ \frac1{2\sqrt{3}} \,a_8 
+\frac{\sqrt{3}}{2}\, a^D_8+\frac{m_2}{\sqrt{2}}=0\,,
\nonumber\\[2mm]
&&\frac12\, a -\frac1{\sqrt{3}}\, a_8 
-\sqrt{3}\;a^D_8+\frac{m_3}{\sqrt{2}}=0 \,.
\label{masslessdyons}
\eeqn

Two remarks are in order here. First and foremost, it is crucially important
to note that the massless dyons 
$D^{11}$ and $D^{22}$ have both electric and magnetic charges $ 1/2$
with respect to the $T^3$ generator of the dual U$(\tN=2)$ gauge group. This means that they
can fill the fundamental representation of this group. Moreover, all dyons 
$D^{lA}$ ($l=1, ... ,\,\tN=2$) can form color doublets.
This is another confirmation 
of the conclusion of Sect.~\ref{secdualgaugegroup}, that
the non-Abelian factor U$(\tN=2)$ of the dual  gauge group gets restored 
in the equal mass limit.

A general  reason ensuring that the dyons $D^{lA}$ ($l=1, ... ,\,\tN$) fill the fundamental
representation of U$(\tN)$ group can be expressed as follows:
due to monodromies the $D^{lA}$ dyons  
pick up magnetic charges of particular  monopoles of SU($N$). 
The magnetic charges of these particular monopoles
are represented by weights rather than roots of the U$(\tN)$ subgroup ($\pm 1/2$ for U$(\tN=2)$,
see (\ref{1moncharge}) and  (\ref{2moncharge})). This is related to the absence of the
AD points associated with collisions of first $\tN$ double roots, 
see (\ref{firstroots}), which,
 of course, is a consequence of the dual theory with the non-Abelian gauge factor U$(\tN)$ being
 not asymptotically free.
Say, in the case of the Abelian dual gauge group ($\tN=0$) studied in Ref.~\cite{SYcross},
the massless dyons pick up integer magnetic charges and, therefore, cannot fill the fundamental
representation of U(2).

The second comment is that the
 dyon charges with respect to each U(1) generator are proportional to each other. This
 guarantees that these dyons are mutually local. Note also, that both the magnetic and electric
charges of the dyon doublet $D^{lA}$ with respect to the $T^8$ generator are $(-1/2) \times$
the charges of the  $D^{33}$ dyon. This is in accord with the result of Ref.~\cite{APS}
where  the charges of the $\tN$-plet with respect to U(1) gauge factors of (\ref{dualgaugegroup})
were shown to be  $(-1/\tN)\times$ the singlet charges.

\subsection{Low-energy effective action}
\label{LEbulkdual}

In this section we present the low-energy theory 
in the $r=3$ vacuum in the domain III, i.e. at small $\xi$ and small $|\Delta m_{PP'}|$ 
(below the AD points). 

As was shown above, the   massless quarks $q^{1A}$ and 
 $q^{2A}$ are transformed into the massless dyons $D^{1A}$ and $D^{2A}$; the latter
 form a fundamental
representation of the dual gauge group U$(\tN=2)$. The $D^{1A}$ and $D^{2A}$ dyons interact with
the U(1) gauge field 
\beq
A_{\mu},
\label{ph}
\eeq
and non-Abelian SU$(\tN=2)$ gauge fields. According to the dyons charges (\ref{dyons}), the third component of 
this SU(2) dual gauge field is the following  linear combination:
\beq
B^3_{\mu}= \frac1{\sqrt{2}}\,(A^{3}_{\mu}+A^{3D}_{\mu})\, .
\label{B3}
\eeq
If the dual gauge group is restored, the $B^{1,2}_{\mu}$ components of the gauge
field  become massless at $m_1=m_2$. Let us check this circumstance. 

The electric and
magnetic charges of the $W$ bosons $B^{1}_{\mu}\mp iB^{2}_{\mu}$ coincide with the charges
of the operators $\tilde{D}_{A2} D^{1A}$ and $\tilde{D}_{A1} D^{2A}$. From (\ref{dyons})
we obtain for the $W$-boson charges
\beq
 B^{1}_{\mu}\mp iB^{2}_{\mu}:\,\,\,
 \left(0,0;\,\pm1,\pm1;\,0,0\right).
\label{B12}
\eeq
These charges determine the mass of these states via the Seiberg--Witten mass formula \cite{SW1}.
We have
\beq
\sqrt{2}|a_3+ a^D_3|=|\Delta m_{12}|,
\label{Wmass}
\eeq
where the first two equations in (\ref{masslessdyons}) are used. We see that this mass tends to
zero at $\Delta m_{12}\to 0$, i.e. $m_1=m_2$, as was expected. Now it is clear that 
the gauge fields (\ref{B3}) and  (\ref{B12}) fill the adjoint
 multiplet $B^{p}_{\mu}$ ($p=1,2,3$) of the non-Abelian  SU(2) factor of the dual gauge group.

Other light states include the  $D^{33}$ dyon and another photon associated with the $T^8$ generator
of the underlying U(3) gauge group broken in the dual theory down to U(2)$\times $U(1),
see Eq.~(\ref{dualgaugegroup}). According to the dyon charges this photon is presented by
the following combination:
\beq
B^8_{\mu}= \frac1{\sqrt{10}}\,(A^{8}_{\mu}+3A^{8D}_{\mu})\, .
\label{B8}
\eeq

In fact, the dyons $D^{lA}\,\,$ ($l=1,2$), $\, D^{33}$ 
and the gauge fields $A_{\mu}$,   $B^{p}_{\mu}$ ($p=1,2,3$),  and $B^8_{\mu}$, together with
their superpartners, 
are the only light states to be included in the 
low-energy effective theory in the domain III. All other states are either
heavy (with masses of the order of $\Lambda$) or decay on 
curves of marginal stability \cite{SW1,SW2,BF,Dorey,svz,SYcross}. In the case at hand CMS
are located 
around the origin in the $\Delta m_{PP'}$ complex planes
and go through the AD point, cf. \cite{svz}. In fact, the $W$ bosons
of the underlying non-Abelian gauge theory, as well as the off-diagonal
states of the quark matrix $q^{kA}$, decay on CMS. We discuss these decay
processes in Sect.~\ref{mmesons}.

Taking this into account we can write the bosonic part of the effective low-energy
action of the theory in the domain III,
\beqn
 S_{III} &=&\int d^4x \left[\frac{1}{4\tilde{g}^2_{2}}
\left(F^{p}_{\mu\nu}\right)^2 
+\frac1{4g^2_1}\left(F_{\mu\nu}\right)^2 +\frac1{4\tilde{g}^2_8}\left(F^8_{\mu\nu}\right)^2
+\frac1{\tilde{g}^2_2}\left|\pt_{\mu}b^p\right|^2 
\right.
\nonumber\\[4mm]
&+& \frac1{g^2_1}
\left|\partial_{\mu}a\right|^2 +\frac1{\tilde{g}^2_8}
\left|\partial_{\mu}b^8\right|^2
+\left|\nabla^1_{\mu}
D^A\right|^2 + \left|\nabla^1_{\mu} \tilde{D}_A\right|^2+
\left|\nabla^2_{\mu}
D^3\right|^2 + \left|\nabla^2_{\mu} \tilde{D}_{3}\right|^2
\nonumber\\[4mm]
&+&
\left. V(D,\tilde{D},b^p,b^8,a)\right]\,,
\label{SIII}
\eeqn
where $b^p$ and $b^8$
are the scalar \ntwo superpartners of gauge fields $B^{p}_{\mu}$ and $B^{8}_{\mu}$,
while $F_{\mu\nu}^p$, $F_{\mu\nu}^8$ are  their field strengths,
\beq
b^3= \frac{1}{\sqrt{2}}\,(a^{3}+a^{3}_D)\;\;\; {\rm for}\;\;\; p=3,
\qquad b^8= \frac{1}{\sqrt{10}}\,(a^{8}+3a^{8}_D).
\label{bbp}
\eeq
Covariant derivatives are defined in accordance
with the charges of the  $D^l$ and $D^3$ dyons. Namely,
\beqn
\nabla^1_\mu & = & 
=\pt_{\mu}-i\left(\frac12 A_{\mu}+\sqrt{2}\,B^p_{\mu}\frac{\tau^p}{2}
 +\frac{\sqrt{10}}{2\sqrt{3}}\,B^8_{\mu}\right)\,,
\nonumber\\[3mm]
\nabla^2_\mu & = & 
=\pt_{\mu}-i\left(\frac12 A_{\mu}
 -\frac{\sqrt{10}}{\sqrt{3}}\,B^8_{\mu}\right)\,.
\label{nablaD}
\eeqn
The coupling constants $g_1$, $\tilde{g}_8$ and $\tilde{g}_2$ 
correspond to two U(1) and the SU(2) gauge groups, respectively.
The scalar potential $V(D,\tilde{D},b^p,b^8,a)$ in the action (\ref{SIII})
is 
\beqn
&& V(D,\tilde{D},b^p,b^8,a) =
 \frac{\tilde{g}^2_2}{4}
\left( \bar{D}_A\tau^p D_A -
\tilde{D}_A \tau^p \bar{\tilde{D}}^A \right)^2
\nonumber\\[3mm]
&+& \frac{10}{3}\frac{g^2_8}{8}
\left(|D^A|^2 -|\tilde{D}_A|^2 -2|D^3|^2 +
2|\tilde{D}_3|^2 \right)^2
\nonumber\\[3mm]
&+& \frac{g^2_1}{8}
\left(|D^A|^2 -|\tilde{D}_A|^2 +|D^3|^2 -
|\tilde{D}_3|^2 
-3 \,\xi\right)^2
\nonumber\\[3mm]
&+& \tilde{g}_2^2\left| \tilde{D}_A \tau^p D_A
\right|^2+
\frac{g^2_1}{2}\left| \tilde{D}_A D^A+
\tilde{D}_3 D_3 \right|^2
+\frac{10}{3}\frac{\tilde{g}_8^2}{2}\left| \tilde{D}_A D^A-
2\tilde{D}_3 D^3\right|^2
\nonumber\\[3mm]
&+&\frac12 \left\{ \left|a+\tau^p\sqrt{2}\,b^p +\sqrt{\frac{10}{3}}\,b^8+\sqrt{2}m_A
\right|^2\left(|D^A|^2+|\tilde{D}_A|^2\right)\right.
\nonumber\\[3mm]
&+&\left.
\left|\;a-2\sqrt{\frac{10}{3}}\,b^8+\sqrt{2}m_3 \;
\right|^2\left(|D^3|^2+|\tilde{D}_3|^2\right) \right\}\,.
\label{potIII}
\eeqn

\vspace{2mm}

Now we are ready move to the desired limit of the equal quark masses, $\Delta m_{PP'}=0$.
The vacuum of the 
theory  (\ref{SIII}) is
located at the following values of the scalars $a$, $b^p$ and $b^8$:
\beq
\langle a\rangle=-\sqrt{2}\,m\,,\qquad \langle b^p \rangle= 0\,,
\qquad\langle b^8 \rangle=0\,,
\label{abvev}
\eeq
while the VEVs of dyons are determined by the FI parameter $\xi$ and can be chosen as
\beqn
\langle D^{lA}\rangle &=&\sqrt{
\xi}\,
\left(
\begin{array}{ccccc}
0 &  0 & 0 & 1 & 0\\
0 &  0 & 0 & 0 & 1\\
\end{array}
\right),
\qquad \langle \bar{\tilde{D}}^{lA}\rangle =0,
\nonumber\\[4mm]
\langle D^{33}\rangle &=& \sqrt{\xi}, \qquad \langle\bar{\tilde{D}}^{33}\rangle =0\,.
\label{Dvev}
\eeqn

In fact, for  the particular choice of quark masses (\ref{masschoice}) we deal with in this section
it is impossible to see which particular flavors of dyons develop VEVs. In the equal mass
limit all $r=3$ isolated vacua coalesce and become a root of the Higgs branch. In Sect.~\ref{split}
we will be able to relax the condition (\ref{masschoice}) and show that, in fact, the $(1,2,3)$ vacuum
we started from at large $\xi$ transforms into the $(4,5,3)$ vacuum of the dual theory 
at small $\xi$, as  shown in (\ref{Dvev}).

Let us calculate the dimension of the Higgs branch which emerges in the equal mass limit.
We have 
\beqn
&& {\rm dim} {\cal H}\,\Big|_{\xi\ll\Lambda}= 4\tN N_f + 4(N-\tN) - 2\tN^2 -\tN^2 -\tN^2
\nonumber\\[2mm]
&& - 2(N-\tN)-(N-\tN)-(N-\tN)=4\tN N\,,
\label{dimHIII}
\eeqn
where we take into account that we have $4\tN N_f + 4(N-\tN)$ dyon real degrees of freedom
and subtract $2\tN^2 + 2(N-\tN)$ $F$-term conditions, $\tN^2 + (N-\tN)$ $D$-term conditions
 and $\tN^2 + (N-\tN)$ phases eaten by the Higgs mechanism, see (\ref{potIII}).

Now we see that the dimension of the Higgs branch  at small $\xi$ coincides
with the dimension of the Higgs branch (\ref{dimH}) at large $\xi$. This strongly supports
our arguments \cite{SYcross}
that we have a crossover transition between two domains I and III, rather than a phase transition.

From Eqs.~(\ref{abvev}) and (\ref{Dvev}) we see that both, the gauge U(2) and flavor SU(5) groups, are
broken in the vacuum. However, the color-flavor locked form of (\ref{Dvev}) guarantees that the diagonal
global SU($\tN=2)_{C+F}$ survives. More exactly, the  unbroken global group of the dual
theory is 
$$
{\rm SU}({3})_F\times  {\rm SU}(2)_{C+F}\times {\rm U}(1)\,.
$$  
For generic 
$N$ and $\tN$ the  unbroken global group of the dual
theory is 
\beq
 {\rm SU}(N)_F\times  {\rm SU}(\tN)_{C+F}\times {\rm U}(1)\,.
\label{c+fd}
\eeq
Here SU$(\tN)_{C+F}$ is a global unbroken color-flavor rotation, which involves the
first $\tN$ flavors, while SU$(N)_F$ factor stands for the flavor rotation of the 
remaining $N$ dyons.
Thus, a color-flavor locking takes place in the dual theory too. Much in the same way as 
in the original microscopic theory, the presence of the global SU$(\tN)_{C+F}$ group 
is the  reason behind formation of the non-Abelian strings.
 For generic quark masses the  global symmetry  (\ref{c+f}) is broken down to 
U(1)$^{N_f-1}$. In parallel with the original microscopic theory, the dimension of the base of the Higgs branch
($2N\tN$) coincides with the number of the broken global generators for the symmetry breaking
pattern (\ref{c+fd}), see (\ref{breakpattern}).

Please, observe that in the equal mass limit 
the global unbroken symmetry (\ref{c+fd}) of the dual theory at small
$\xi$ coincides with the global group (\ref{c+f}) present in the
$r=N$ vacuum of the original microscopic theory at large
$\xi$. This is, of course, expected and presents a  check of our results. Note
however, that this global symmetry is realized in two distinct ways in two dual theories.
As was already mentioned, the quarks and U($N$) gauge bosons of the original theory at large $\xi$
come in the $(1,1)$, $(N^2-1,1)$, $(\bar{N},\tN)$, and $(N,\bar{\tN})$
representations of the global group (\ref{c+f}), while the dyons and U($\tN$) gauge 
bosons form $(1,1)$, $(1,\tN^2-1)$, $(N,\bar{\tN})$, and 
$(\bar{N},\tN)$ representations of (\ref{c+fd}). We see that adjoint representations of the $(C+F)$
subgroup are different in two theories. A similar phenomenon was detected in \cite{SYcross}
for the Abelian dual theory in the case $\tN=0$.

 We traced the evolution of light quarks
from the domain I to  II and then back to the equal mass limit along the Coulomb branch at zero $\xi$. We demonstrated 
that quarks transform into dyons along the way,
picking up magnetic charges. For  consistency of
our analysis it is instructive to
consider another route from the domain I to the domain III, namely the one along the line
 $\Delta m_{AB}=0$. On this 
line we keep the global group (\ref{c+fd}) unbroken. Then we obtain a surprising result:
the quarks and gauge bosons
which form the  adjoint $(N^2-1)$ representation  of SU($N$) at large $\xi$ and the dyons and gauge bosons which form the  adjoint $(\tN^2-1)$ representation  of SU($\tN$) at small $\xi$ are, in fact, {\em distinct} states.
How can this occur?

Since we have a crossover  between the domains I and III rather than  a phase 
transition,
this means that in the full microscopic theory the $(N^2-1)$  adjoints of SU($N$) become heavy 
and decouple as we pass from the
domain I to III along the line $\Delta m_{AB}=0$. Moreover, some 
composite $(\tN^2-1)$ adjoints  of SU($\tN$), which are 
heavy  and invisible in the low-energy description in the domain I become light in 
the  domain III and form the $D^{lK}$ dyons
 ($K=N+1,...,N_f$) and gauge bosons $B^p_{\mu}$. The phenomenon of level crossing
 takes place. Although this crossover is smooth in the full theory,
from the standpoint of the low-energy description the passage from  the domain I to 
 III means a dramatic change: the low-energy theories in these domains are 
completely
different; in particular, the degrees of freedom in these theories are different.

This logic leads us to the following conclusion. In addition to light dyons and gauge bosons 
 included in  the low-energy theory (\ref{SIII}), in the domain III at small $\xi$ we have
heavy  fields (with masses of the order of $\Lambda$) which form the adjoint representation
$(N^2-1,1)$ of the global symmetry (\ref{c+fd}). These are screened (former)  quarks 
and gauge bosons from the domain I continued into III.
 Let us denote them as $M_P^{P'}$ ($P,P'=1,...,N$). In Sect.~\ref{mmesons} we will 
 discuss them in more detail and reveal
their physical nature  in the domain III.

By the same token, it is  seen that in the domain I, in addition to the light quarks and gauge bosons,
we  have heavy fields $M_K^{K'}$ ($K,K'=N+1,...,N_f$), which  form the  adjoint $(\tN^2-1)$ representation  of SU($\tN$).
This is schematically depicted in Fig.~\ref{figevol}.

\begin{figure}
\epsfxsize=7cm
\centerline{\epsfbox{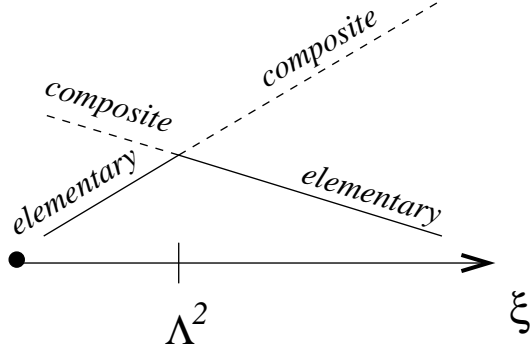}}
\caption{\small Evolution of the SU$(N)$ and SU$(\tN)$ $W$ bosons vs. $\xi$. 
On both sides of the level crossing at $\xi=\Lambda^2$ the global groups are SU$(N)\times$SU$(\tN)$,
however, above $\Lambda^2$ it is SU$(N)_{C+F}\times$SU$(\tN)_F$ while
below $\Lambda^2$ it is SU$(N)_F\times$SU$(\tN)_{C+F}$.}
\label{figevol}
\end{figure}

It is quite plausible to suggest that these fields $M_P^{P'}$ and $M_K^{K'}$ are Seiberg's mesonic fields
\cite{Seiberg:1994pq,IS},
which occur in the dual theory upon breaking of \ntwo supersymmetry by the mass-term
 superpotential $\mu[A^2 +(A^a)^2]$ for the adjoint fields when we take the limit 
$\mu\to\infty$. In this limit  our theory becomes \none QCD. In the \ntwo limit the $M_P^{P'}$ and $M_K^{K'}$
fields  are heavy, with
masses $\sim \Lambda$,  and 
are absent in the low-energy action (\ref{SIII}). 
However, in the $\mu\to\infty$ limit it is the  
  \none scale $\Lambda_{{\cal N}=1}$ that is  fixed, $$\Lambda_{{\cal N}=1}^{2N-\tN}=
\mu^{N}\Lambda^{N-\tN}\,,$$ implying that $\Lambda\to 0$. The $M_{AB}$ fields  might
become light in the limit of \none QCD. Previously, these $M_{AB}$ fields were not identified 
in the \ntwo theory.

\section{Confined monopoles}
\label{conf}
\setcounter{equation}{0}

Since the quarks are in the Higgs regime in the 
original microscopic theory, the monopoles are confined. It is known
\cite{T,SYmon,HT2} that when we introduce a nonvanishing FI parameter $\xi$  
in  \ntwo QCD  with the gauge group U$(N)$, we confine
the 't Hooft--Polyakov monopoles of the SU$(N)$ subgroup to  the string.
In fact, they become string junctions of two elementary non-Abelian strings.
They are seen as kinks in the world-sheet theory (\ref{wcp}) at large $\xi$, and 
as kinks in the dual world-sheet theory 
in the domain III at small $\xi$ (see Sect.~\ref{2Ddual}). In this domain it is dyons, rather than quarks,
that  condense. Therefore, here we deal with oblique confinement \cite{thooft}.

In this section we will determine the elementary string fluxes  in the classical limit 
in the domain III and show that the elementary monopole
fluxes can be absorbed by two strings. Hence,
the monopoles are indeed represented by junctions of different strings.

As a warm up example,  we start from reviewing  matching of the
monopole and strings fluxes in the domain I at large $\xi$.
To this end we go to the quasiclassical limit in the world-sheet theory (\ref{wcp}), i.e.  $\Delta m_{PP'}\gg\Lambda$,
where the non-Abelian strings become Abelian $Z_N$ strings, 
see Ref.~\cite{SYrev} for more details. 

As in Sect.~\ref{monodro}, we restrict ourselves to the simplest example $N=3$, $\tN=2$.
Consider one of three $Z_3$ strings which occur due to winding of the $q^{11}$ quark  at infinity,
\beq
q^{11}(r\to\infty) \sim\sqrt{\xi}\,e^{i\alpha},
\qquad q^{22}(r\to\infty) \sim q^{33}(r\to\infty)\sim \sqrt{\xi},
\label{qwind}
\eeq
see (\ref{qvev}). Here $r$ and $\alpha$ are the polar coordinates in the plane $i=1,2$ orthogonal 
to the string axis.
This implies the following behavior of the gauge potentials at $r\to\infty$:
\beqn
&& \frac12 A_i +\frac12 A_i^3 + \frac1{2\sqrt{3}} A_i^8 \sim \pt_i \alpha\,,
\nonumber\\[2mm]
&& \frac12 A_i -\frac12 A_i^3 + \frac1{2\sqrt{3}} A_i^8 \sim 0\,,
\nonumber\\[2mm]
&& \frac12 A_i  - \frac1{\sqrt{3}} A_i^8 \sim 0\,,
\eeqn
see the quark charges in (\ref{quarkcharges}). The solution to these equations is
\beqn
&&  A_i  \sim \frac23 \,\pt_i \alpha\,,
\nonumber\\[2mm]
&& A_i^3  \sim \pt_i \alpha\,,
\nonumber\\[2mm]
&&   A_i^8 \sim \frac1{\sqrt{3}}\,\pt_i \alpha\,.
\label{gaugewindI}
\eeqn
It determines the string gauge  fluxes $\int dx_i A_i$, $\int dx_i A^3_i$ and $\int dx_i A^8_i$, respectively.
The integration above is performed over a large circle in the $(1,2)$ plane.
Let us call this string $S_1$.

Next, we define the string charges as 
\beq
\int dx_i (A_i,\,A^D_i;\,A^3_i,\,A^{3D}_i;\,A^8_i,\,A^{8D}_i)= 4\pi\,(n_e,\,n_m;\,n^3_e,\,n^3_m;\,n^8_e,\,n^8_m)\,.
\label{defstrch}
\eeq
This definition ensures that the string has the same charge as a trial monopole which can be attached to
the string endpoint. In other words, the flux of the given  string is the flux of the trial monopole\,\footnote{This trial monopole does not necessarily exist in our
theory. In fact, in U($N$) theories we deal with here, the strings are stable and there are no monopoles
in the theory {\em per se} which could break these strings. The SU$(N)$ monopoles are rather string junctions, so they are 
attached to {\em two} strings, as we will see below.} 
sitting on string's end, with
the charge defined by (\ref{defstrch}).

In particular, according to this definition, the charge of the string with the fluxes (\ref{gaugewindI})
is 
\beq
\vec{n}_{S_1}=\left(0,\,\frac13;\,0,\,\frac12;\,\,0,\,\frac1{2\sqrt{3}}\right).
\label{S1}
\eeq
Since this string is formed through the quark condensation, it is  magnetic;   its 
 charges with respect to the Cartan subalgebra of the SU(3) group
are represented by the weight vector, as seen from (\ref{S1}). 

There are other two elementary strings $S_2$ and $S_3$ in U(3) which arise due to winding of 
$q^{22}$ and 
$q^{33}$
quarks, respectively. Repeating the above procedure for these strings we get their charges,
\beq
\vec{n}_{S_2}=\left(0,\,\frac13;\,\,0,\,-\frac12;\,\,0,\,\frac1{2\sqrt{3}}\right),\quad
\vec{n}_{S_3}=\left(0,\,\frac13;\,0,\,0;\,0,\,-\frac1{\sqrt{3}}\right).
\label{S23}
\eeq
It is easy to check that each of three elementary SU(3) monopoles is confined by two elementary
strings.  Consider, say, the monopole with the charge $(0,\,0;\,0,\,1;\,0,\,0)$. This charge can be written as
a difference of the charges of two elementary strings $S_1$ and $S_2$, namely, 
\beq
(0,\,0;\,0,\,1;\,0,\,0)= \vec{n}_{S_1}-\vec{n}_{S_2}\,.
\label{confI}
\eeq
This means that this monopole  is a junction of these two strings at large $\xi$, with $S_1$ string having
the outgoing flux while
$S_2$ the  incoming flux.

Now we are ready to turn to the monopole confinement in the domain III,  described
by the dual theory (\ref{SIII}). Consider the $\tilde{S}_1$ string
arising due to winding of the  $D^{14}$ dyon. At $r\to\infty$ we have 
\beq
D^{14}(r\to\infty) \sim\sqrt{\xi}\,e^{i\alpha},
\qquad D^{25}(r\to\infty) \sim D^{33}(r\to\infty)\sim \sqrt{\xi},
\label{Dwind}
\eeq
see (\ref{Dvev}).
Taking into account the dyon charges quoted in Eq.~(\ref{dyons}) (the $D^{14}$ and $D^{25}$ dyons   have the same 
electric and magnetic charges as 
$D^{11}$ and $D^{22}$, respectively)
we derive the behavior of the gauge potentials at infinity,
\beqn
&& \frac12 A_i +\frac12 A_i^3 +\frac12 A_i^{3D}+ \frac1{2\sqrt{3}} A_i^8 
+ \frac{\sqrt{3}}{2} A_i^{8D}\sim \pt_i \alpha\,,
\nonumber\\[2mm]
&& \frac12 A_i -\frac12 A_i^3 -\frac12 A_i^{3D}+ \frac1{2\sqrt{3}} A_i^8 
+ \frac{\sqrt{3}}{2} A_i^{8D} \sim 0\,,
\nonumber\\[2mm]
&& \frac12 A_i  - \frac1{\sqrt{3}} A_i^8 - \sqrt{3} A_i^{8D}\sim 0\,,
\eeqn
which, in turn, implies
\beqn
&&  A_i  \sim \frac23 \,\pt_i \alpha\,,
\nonumber\\[2mm]
&& \frac12 A_i^3 +\frac12 A_i^{3D}  \sim \frac12 \,\pt_i \alpha\,,
\nonumber\\[2mm]
&&   \frac1{2\sqrt{3}} A_i^8 
+ \frac{\sqrt{3}}{2} A_i^{8D} \sim \frac1{6}\,\pt_i \alpha \,.
\label{gaugewindint}
\eeqn
The combinations orthogonal to those which appear in (\ref{gaugewindint})
are required to tend to zero at infinity, 
namely, $A_i^3 - A_i^{3D}\sim 0$ and $A_i^{8D} -3A_i^{8} \sim 0$.
As a result we get 
\beqn
&&  A_i  \sim \frac23 \,\pt_i \alpha \,,\qquad A_i^D  \sim 0\,,
\nonumber\\[2mm]
&& A_i^3  \sim \frac12 \,\pt_i \alpha\,, \qquad A_i^{3D}  \sim \frac12 \,\pt_i \alpha\,,
\nonumber\\[2mm]
&&   A_i^8 \sim \frac1{10\sqrt{3}}\,\pt_i \alpha \,, \qquad A_i^{8D} \sim \frac{\sqrt{3}}{10}\,\pt_i \alpha\,.
\label{gaugewindIII}
\eeqn
These expressions determine the charge of the $\tilde{S}_1$ string,
\beq
\vec{n}_{\tilde{S}_1}=\left(0,\,\frac13;\,-\frac14,\,\frac14;\,-\frac{\sqrt{3}}{20},\,\frac1{20\sqrt{3}}\right).
\label{tS1}
\eeq

Paralleling the above analysis
we determine the charges of two other $Z_3$ strings which are due to windings of 
$D^{25}$ and $D^{33}$, respectively. We get
\beq
\vec{n}_{\tilde{S}_2}=\left(0,\,\frac13;\,\frac14,\,-\frac14;\,-\frac{\sqrt{3}}{20},\,\frac1{20\sqrt{3}}\right),
\quad
\vec{n}_{\tilde{S}_3}=\left(0,\,\frac13;\,0,\,0;\,\frac{\sqrt{3}}{10},\,-\frac1{10\sqrt{3}}\right).
\label{tS23}
\eeq
Now we can check that each of three SU(3) monopoles can be confined by two strings. Say, for 
the monopole with the  charge $(0,0;\,0,1;\,0,0)$ we have 
\beq
(0,\,0;\,0,\,1;\,0,\,0)= (\vec{n}_{\tilde{S}_1}-\vec{n}_{\tilde{S}_2} )+\frac12\left(
\vec{n}_{D^{14}}-\vec{n}_{D^{25}}\right),
\label{confIII}
\eeq
where $\vec{n}_{D^{14}}$ and $\vec{n}_{D^{25}}$ are the charges of the $D^{14}$ and $D^{25}$ dyons
given in (\ref{dyons}).
Only a part of the monopole flux is confined to the strings. The remainder  of its flux is
screened by the condensate of the $D^{14}$ and $D^{25}$ dyons. In a similar manner we can check confinement of
the other two SU(3) monopoles.

We see that although the quark charges change as we pass from the domain I to III, and they 
become dyons, this does {\em not} happen with the monopoles. The
monopole states do not change their charges. They are confined
in both domains I and III, being junctions of two different elementary strings. In the domain
III in the dual theory 
there is a peculiarity: not all of the monopole flux is carried by  two attached strings; 
a part of it is screened by dyon condensate.

Our result provides an explicit counterexample to the commonly
accepted belief that if monopoles are confined in the original theory,
then it is  quarks that are confined in the
dual theory. Above we demonstrated that monopoles rather than quarks are  confined in the domain III.
The failure of this folklore  belief eliminates a paradox
mentioned in \cite{Shifman:2007kd} where this folklore was tacitly assumed.

We can check that the dyons whose charges are the sum of the monopole and $W$-boson
 charges are also confined. As an example of such a state it is worth considering the dyon with the charge
$(0,\,0;\,1,\,1;\,0,\,0)$ in the domain II. Below the crossover, in the domain III, its charge is
shifted by the monopole charge due to monodromy. In the domain III this dyon has the charge
$(0,\,0;\,1,2;\,0,\,0)$. Therefore, we have
\beq
(0,0;\,1,2;\,0,0)= (\vec{n}_{\tilde{S}_1}-\vec{n}_{\tilde{S}_2} )+\frac32\left(
\vec{n}_{D^{14}}-\vec{n}_{D^{25}}\right),
\label{confdIII}
\eeq
which shows that this dyon is confined by two strings, $\tilde{S}_1$ and $\tilde{S}_2$,
while the remainder of its flux is screened by condensation of the  $D^{14}$ and $D^{25}$ dyons.

\section{Splitting the quark masses}
\label{split}
\setcounter{equation}{0}

In this section we relax the condition (\ref{masschoice}) and split the masses of the first $\tN$
quarks (out of $N$ quarks which develop VEVs at large $\xi$) and $\tN$ extra quarks. If all
masses are generic, the Higgs branch disappears, and we have $C_{N_f}^N$ isolated $r=N$ vacua in the 
original theory (\ref{model}) at large $\xi$ in the domain I. Again,  we  consider one of these vacua,
namely, the $(1,...,N)$ vacuum. We will show that in the domain III at small $\xi$ it converts\,\footnote{Of course,  the total number of  vacua in the dual theory ($C_{N_f}^{\tN}$
with generic masses) matches the number of  vacua in the original theory, $C_{N_f}^N=C_{N_f}^{\tN}$.} into
the $(N+1, ... ,\, N_f,\, \tN+1,...,N)$ vacuum, as indicated in (\ref{Dvev}) for the case $N=3$ and $\tN=2$.

If the condition (\ref{masschoice}) is fulfilled the dual theory (\ref{SIII}) is IR rather  then asymptotically
 free. Once we relax this condition, it becomes asymptotically free at the scales below $\Delta m_{PK}$
($P=1,...,N$ and $K=N+1,...,N_f$).  We assume that all mass differences $\Delta m_{PK}$ are of the
the same order. In fact, the theory generates its own low-energy scale
\beq
\tilde{\Lambda}_{\rm le}^{\tN}\, =\,\, \frac{\Delta m_{PK}^N}{\Lambda^{N-\tN}}\, .\rule{0mm}{10mm}
\label{tL}
\eeq

\vspace{2mm}

\noindent
In order to guarantee the week coupling regime in the dual theory (\ref{SIII})  we cannot choose  $\xi$ too small
in the domain III. We have we assume that
\beq
\tilde{\Lambda}_{\rm le}\ll\sqrt{\xi}\,.
\label{tLxi}
\eeq
Since $\Lambda\gg\sqrt{\xi}$ in the domain III the above condition requires, in turn, that 
the mass splittings
 $\Delta m_{PK}$ not to be too large. We impose the following constraint: 
\beq
\Delta m_{PK}\ll\Lambda\,,\quad P=1, ... ,\, N\,,\quad K=N+1, ... ,\, N_f\,.
\label{DeltamL}
\eeq

In parallel with our discussion in Sect.~\ref{bulkdual} we pass from the domain I to  II at weak coupling
and then  to the domain III along the Coulomb branch (at $\xi=0$), using the Seiberg--Witten exact solution
of the theory.
The role of the $\Delta m$ variable   in Fig.~\ref{figphasediag} is played by the mass differences $\Delta m_{PP'}$ and 
$\Delta m_{KK'}$ which we assume to be of the same order.

\subsection{The Seiberg--Witten curve}
\label{swic}

To make our discussion simpler in this section   we  again consider the  example
of the U(3) gauge theory with $N_f=5$. At first, we relax just the first of the conditions (\ref{masschu3})
and define
\beq
\Delta m_{14}\equiv m_1-m_4\,,\qquad m_{14}\equiv \frac12(m_1+m_4)\,,
\eeq
keeping $m_5=m_2$. Then the Seiberg--Witten curve takes the form 
\beqn
 y^2
 &=&
 \left(x+\frac{\m_2}{\sqrt{2}}\right)^2 \left[(x-\phi_1)^2(x-\phi_3)^2 
-4\frac{\Lambda}{\sqrt{2}}
\left(x+\frac{\m_{14}+\Delta m_{14}}{\sqrt{2}}\right)
\right.
\nonumber\\[4mm]
&\times& 
\left. \left(x+\frac{\m_{14}-\Delta m_{14}}{\sqrt{2}}
\right)\left(x+\frac{\m_5}{\sqrt{2}}\right)\right],
\label{U3curvesplit}
\eeqn
where we substituted the solution (\ref{phi12}) for $\phi_2$. The double root $e_3=e_4$ is given
in the second equation in (\ref{roots1-4}).
Next we parametrize
\beq
\phi_1 = -\frac{\m_{14}}{\sqrt{2}} + \chi \,,
\label{chidef}
\eeq
where $\chi$ is  small. Also we
 shift $x$,
\beq
x=-\frac{\m_{14}}{\sqrt{2}} +z\,,
\label{xshiftsp}
\eeq
and arrive at
\beqn
 y^2
 &=&
   \left(x+\frac{\m_2}{\sqrt{2}}\right)^2 
\left[z^2\left(z+\frac{\Delta m_{31}}{\sqrt{2}} -\frac{\Lambda}{\sqrt{2}}\right)^2
+(-2\chi z +\chi^2)
\right.
\nonumber\\[4mm]
&\times& 
\left(z+\frac{\Delta m_{31}}{\sqrt{2}} +\frac{\Lambda}{\sqrt{2}}\right)^2
\left.
+4\frac{\Lambda}{\sqrt{2}}\, \frac{\Delta m_{14}^2}{8} \left(z+\frac{\Delta m_{31}}{\sqrt{2}}\right)\right],
\label{zcurve}
\eeqn
where $\Delta m_{31}=m_3-m_{14}$.  Here we use (approximately) the  solution (\ref{phi3}) for $\phi_3$
obtained for unsplit masses.

Next, we look for roots of (\ref{zcurve}) located near the unperturbed values of $e_1$ and $e_2$ 
(see Eq.~(\ref{roots1-4})), so that $z$
is close to zero. The curve (\ref{zcurve}) approximately gives a quadratic
equation for these roots,
\beq
z^2-(2\chi\,z -\delta^2)\left(\frac{\Delta m_{31} +\Lambda}
{\Delta m_{31} -\Lambda}\right)^2 +
4\Lambda\, \frac{\Delta m_{14}^2}{8} \,\frac{\Delta m_{31}}
{\left(\Delta m_{31} -\Lambda\right)^2}=0\,.
\label{queqn}
\eeq
We need to find such $\chi$ that ensures that the two roots of this equation coincide.
This is an easy exercise leading to 
\beq
\chi =  \pm\frac{\Delta m_{14}}{2\sqrt{2}}\,\left(\frac{\Delta m_{31} -\Lambda}
{\Delta m_{31} +\Lambda}\right),
\label{chi}
\eeq
which gives, in turn,
\beq
\phi_1 = -\frac{\m_{14}}{\sqrt{2}} -\frac{\Delta m_{14}}{2\sqrt{2}}\,\left(\frac{\Delta m_{31} -\Lambda}
{\Delta m_{31} +\Lambda}\right).
\label{phi1}
\eeq
The corrected roots $e_1$ and $e_2$ are
\beq
e_1 =e_2= -\frac{\m_{14}}{\sqrt{2}} -\frac{\Delta m_{14}}{2\sqrt{2}}\,\left(\frac{\Delta m_{31} +\Lambda}
{\Delta m_{31} -\Lambda}\right).
\label{e12}
\eeq
Here we pick up only the solution with the  minus sign for $\chi$ in (\ref{chi}). The reason is that
in the quasiclassical regime of large $\Delta m_{31}$ ($\Delta m_{31}\gg\Lambda$) the solution
(\ref{phi1}) is determined  by $m_1$, see Eq.~(\ref{classphi}). This corresponds to the $(1,2,3)$ vacuum
we started from  in the domain I and II. The opposite sign would correspond to the $ (4,2,3)$ vacuum.

Please, observe that 
\beq
\phi_1=
\left\{
\begin{array}{l}\rule{0mm}{5mm}
-\frac{\m_{1}}{\sqrt{2}},\qquad  |m_{31}|\gg\Lambda\,,
\\[4mm]
 -\frac{\m_{4}}{\sqrt{2}},\qquad   |m_{31}|\ll\Lambda\,.
\end{array}
\right.
\label{phi1lim}
\eeq
We see that $\phi_1$ evolves from $m_1$ to $m_4$ as we reduce $\Delta m_{31}$ moving from the domain
II towards III and then inside III.
By the same token, we can split the $m_2$ and $m_5$ masses  and study the behavior of $\phi_2$.
In this way we get 
\beq
\phi_2=
\left\{
\begin{array}{l}\rule{0mm}{5mm}
-\frac{\m_{2}}{\sqrt{2}},\qquad  |m_{32}|\gg\Lambda\,,
\\[4mm]
 -\frac{\m_{5}}{\sqrt{2}},\qquad   |m_{32}|\ll\Lambda\,.
\end{array}
\right.
\label{phi2lim}
\eeq

These results demonstrate that the $(1,2,3)$ vacuum  of  the original theory (\ref{model})
in the domains I and II converts into the $(4,5,3)$ vacuum  of the dual theory (\ref{SIII}) as we
go deep into the domain III,
\beq
(1,2,3)\Big|_{\rm I,II} \to (4,5,3)\Big|_{\rm III}\,,
\eeq
or, in the case of generic $N$ and $\tN$,
\beq
(1, ... ,\, N)\Big|_{\rm I,II} \to (N+1, ... , \,N_f,\,\,\tN+1, ... ,\, N)\Big|_{\rm III}\,.
\label{jump}
\eeq
In other words, if we pick up the vacuum (\ref{avev}), (\ref{qvev})
in our  theory (\ref{model}) at large $\xi$ in the domain I and reduce $\xi$ passing to the domain III,
the system goes through a crossover transition and ends up in the vacuum of the dual theory (\ref{SIII}) with
the following VEVs of the adjoint scalars:
\beqn
&& \left\langle \frac12\, a + \frac{\tau^p}{2}\, \sqrt{2}\,b^a+\frac12\sqrt{\frac{10}{3}}\,b^8
\right\rangle = - \frac1{\sqrt{2}}
 \left(
\begin{array}{cc}
m_{4} &  0 \\
0 &  m_{5}\\
\end{array}
\right), 
\nonumber\\[3mm]
&& \left\langle \frac12\, a -\sqrt{\frac{10}{3}}\,b^8\right\rangle 
=- \frac1{\sqrt{2}}\,m_3,
\label{abvevsp}
\eeqn
while the VEVs of dyons are given in Eq.~(\ref{Dvev}), where Eqs.~({\ref{abvevsp}) and (\ref{Dvev})
are specified for $N=3$, $\tN=2$.

Equation (\ref{abvevsp}) ensures that the conditions for  the massless dyons (\ref{masslessdyons})
are modified when $m_1\neq m_4$ and $m_2\neq m_5$ as follows:
\beqn
&&\frac12 \,a+ \frac12 \,a_3+ \frac12 \,a^D_3+ \frac1{2\sqrt{3}}\, a_8
+\frac{\sqrt{3}}{2}\, a^D_8 +\frac{m_4}{\sqrt{2}}=0\,,
\nonumber\\[2mm]
&&\frac12 \,a- \frac12 \,a_3 - \frac12 \,a^D_3+ \frac1{2\sqrt{3}} \,a_8 
+\frac{\sqrt{3}}{2}\, a^D_8+\frac{m_5}{\sqrt{2}}=0\,,
\nonumber\\[2mm]
&&\frac12\, a -\frac1{\sqrt{3}}\, a_8 
-\sqrt{3}\;a^D_8+\frac{m_3}{\sqrt{2}}=0 \,.
\label{masslessdyonssp}
\eeqn

We pause here to make one 
last comment. The pole present in (\ref{phi1}) at 
$\Delta m_{31} +\sqrt{2}\Lambda=0$   has no physical meaning. It is canceled out in
the  expressions for the standard coordinates on the Coulomb branch 
$$u_k= \phi_1^k+\phi_2^k+\phi_3^k\,.$$ To see that this is indeed the case
one has to consider small deviations of $\phi_3$ from its approximate solution (\ref{phi3}).

\subsection{The \boldmath{$W$}-boson mass}
\label{wbosmas}

In this section we will present another argument supporting our claim
 that as one passes through the crossover, the vacuum we had in the domain III
 turns into a distinct $r=N$ vacuum, as
shown in Eq.~(\ref{jump}). 

Consider again the already familiar 
example with $N=3$ and  $\tN=2$. On the Coulomb branch in the $(1,2,3)$ vacuum 
at weak coupling (in the domain II at $\xi=0$) the mass of the $A_{\mu}^{1,2}$ gauge fields  is 
\beq
m_W\Big|_{\rm II}=\sqrt{2}|a_3|=\left|\Delta m_{12}\right|\,.
\label{WmassII}
\eeq
Below the crossover, in the domain III, the charged components of the dual SU(2) gauge multiplet 
 are the $B_{\mu}^{1,2}$ fields defined in (\ref{B12}).
In Sect.~3.3 we calculated the mass of these fields
(the $W$-boson mass) in the limit of unsplit quark masses (\ref{masschu3}), see (\ref{Wmass}).
In the  limit (\ref{masschu3}) the $W$-boson mass coincides with the value (\ref{WmassII}). Now 
we will 
split quark masses and show that the $W$-boson mass experiences a jump as we 
pass from the domain II to  III.

Taking into account the charges of the $B_{\mu}^{1,2}$ fields 
--- these fields will be referred to as the $W^{*}$ bosons ---
quoted in Eq.~(\ref{B12}) we arrive at the following expression
for the $W^{*}$-boson masses in the domain III 
at $\xi=0$:
\beq
m_{W^{*}}\Big|_{\rm III}=\sqrt{2}\,\left|a_3+ a^D_3\right|=\,\left|\Delta m_{45}\right|\,.
\label{WmassIII}
\eeq
To derive (\ref{WmassIII})
 we take the difference of two   first  equations in
 (\ref{masslessdyonssp}). Note that both Eqs.~(\ref{WmassII}) and (\ref{WmassIII})
 are exact. We see that, according to (\ref{jump}), the $W$-boson mass experiences  a jump.

It is instructive to check this result by explicit calculation via the Seiberg--Witten curve. The mass of the 
SU(2) $W$ boson coincides with the discontinuity of the following period integral:
\beq
m_W=\frac{\sqrt{2}}{2\pi}\,\left|\Delta\left[\sum_{P=1}^{N}\int_{e_3=e_4}^{e_1=e_2}
\frac{xdx}{x+\frac{m_P}{\sqrt{2}}}-\sum_{K=\tN+1}^{N_f}\int_{e_3=e_4}^{e_1=e_2}
\frac{xdx}{x+\frac{m_K}{\sqrt{2}}}\right]\right|,
\label{periodint}
\eeq
where $\Delta$ means taking the discontinuity of the logarithmic function. Substituting here
the expressions (\ref{e12}) for the $e_1=e_2$ roots  and similar expression for the $e_3=e_4$ roots 
\beq
e_3 =e_4= -\frac{\m_{25}}{\sqrt{2}} -\frac{\Delta m_{25}}{2\sqrt{2}}\,\left(\frac{\Delta m_{32} +\Lambda}
{\Delta m_{32} -\Lambda}\right)
\label{e34}
\eeq
we obtain,  with  logarithmic accuracy,
\beq
m_W\Big|_{\rm II} = \frac{1}{2\pi}\,\left|\Delta\left\{\Delta m_{12}\,\ln{\frac{\Delta m}{\Lambda}}\right\}\right|
\eeq
at large $\Delta m$ ($ \Delta m\gg\Lambda$ where $\Delta m\equiv \Delta m_{31}\sim \Delta m_{32}$).
On the other hand, at small $\Delta m $ ($\Delta m\ll\Lambda$)
\beq
m_{W^{*}}\Big|_{\rm III} = \frac{1}{2\pi}\,\left|\Delta\left\{\Delta m_{45}\,\ln{\frac{\Lambda}{\Delta m}}\right\}\right|\,.
\eeq
Taking the discontinuity of logarithms  we fully confirm 
the results presented in (\ref{WmassII}) and (\ref{WmassIII}).

\vspace{2mm}

The key point of this calculation is Eq. (\ref{e12}) for the $e_1=e_2$ roots 
and the companion expression
(\ref{e34})   for the $e_3=e_4$ roots. Say, the 
double root $e_1=e_2$ tends to $\m_1$ at $\Delta m\gg\Lambda$ and to
$\m_4$ at $\Delta m\ll\Lambda$. A more careful study of the integral in (\ref{periodint}) shows
that the two jumps occur precisely at two AD points (\ref{AD1}) and (\ref{AD2}).

Does the jump of the $W$-boson masses means that the {\em physical spectrum} has 
a genuine discontinuity at the AD points?
Of course, not. 

No real physical phase transitions are implied at these points. The physical spectrum
is continuous. The apparent jump of the $W$-boson mass means that,  in actuality, we have {\em two} 
$W$-boson-like states. Let us denote them as $W$ and $W^{*}$, respectively. 
They have the same electric and magnetic charges ( $(0,0;\pm 1,0;0,0)$ above the 
crossover, and $(0,0;\pm 1,\pm 1;0,0)$ below the crossover, see (\ref{B12})), but distinct global
flavor U(1) charges. Note, that the global group (\ref{c+f}) (or the dual global group (\ref{c+fd})) is broken
by mass differences down to
\beq
U(1)^{N_f-1}\,.
\label{brokengroup}
\eeq
All massive BPS states have nonvanishing charges with respect to this group.
The $W$ bosons acquire nonvanishing global charges due to the color-flavor locking.

Above the crossover (i.e. at large $|\Delta m|$) the $W$ boson has mass (\ref{WmassII}) while
that of $W^{*}$ is 
\beq
m_W^{*}\Big|_{\rm II}=\sqrt{2}
\,\left|a_3 +\frac{\Delta m_{14}}{\sqrt{2}}-\frac{\Delta m_{25}}{\sqrt{2}}\right|= \left|\Delta m_{45}\right|.
\label{WemassII}
\eeq
Below the crossover (i.e. at small $|\Delta m|$) the mass of the $W^{*}$ boson  is given by (\ref{WmassIII}) while that
of $W$ is
\beq
m_W\Big|_{\rm III}=\sqrt{2}
\,\left|a_3+ a^D_3-\frac{\Delta m_{14}}{\sqrt{2}}+\frac{\Delta m_{25}}{\sqrt{2}}\right|=
\left|\Delta m_{12}\right|.
\label{WemassIII}
\eeq
We see that given two states, $W$ and $W^{*}$, the physical spectrum {\em is}  continuous, indeed. 

\section{More on particles in the adjoint representations
of \boldmath{${\rm SU}(N)$} and \boldmath{${\rm SU}(\tN)$}: crossing the boundaries}
\label{mmesons}
\setcounter{equation}{0}

The problem of stability of massive BPS states on the Coulomb branch of our theory (i.e. at $\xi=0$)  needs
additional studies. This is left for future work. Here we will make a few general comments
following from  consistency of our picture. 

It is well known that the $W$
bosons usually do not exist as localized states 
 in the strong coupling regime on the Coulomb branch (speaking in jargon, they ``decay"). They split
into antimonopoles and dyons on CMS on  which   the AD points lie \cite{SW1,BF}.

In our theory this ``decay" involves two steps. Consider the $W$-boson associated with the
$T^3$  generator ($T^3$ $W$ boson for short) with the charge
$(0,0;\,1,0;\,0,0)$ in the domain II.
As we approach the first AD point (\ref{AD1})  from the domain II, the  $T^3$ $W$ boson ``emits"
massless antimonopole with the charge opposite to the one in Eq.~(\ref{1moncharge}).
After we pass by the second AD point
(\ref{AD2}) it ``emits"  massless monopole with the charge (\ref{2moncharge}). 
The net effect is the ``decay"
of the $W$ boson into the $T^3$ antimonopole and 
dyon with the charges $(0,0;\, 0,-1;\,0,0)$ and $(0,0;\,1,1;\,0,0)$, respectively. 
It means that the $W$ boson is absent 
 in the domain III,  in full accord with the analysis of
the SU(2) theory in \cite{BF}.

In our theory we have another $T^3$ $W$-boson-like state, namely, $W^{*}$.
Clearly this state also can ``decay"
 in the same $T^3$ antimonopole and a different dyon\,\footnote{This dyon has the same electric and magnetic charges ($(0,0;\,1,1;\,0,0)$ in the domain II and  the charge 
$(0,0;\,1,2;\,0,0)$ in the domain III)
as   the dyon
associated with the $W$ state, but different global U(1) charges with respect to
(\ref{brokengroup}).} 
as we pass through the crossover.
 In the domain III the $W^{*}$ state   plays the
role of the gauge field of the dual theory. Therefore, we expect that it is stable in the domain III
and ``decays" in the domain II.

This picture is valid on the Coulomb branch at $\xi=0$.  As we switch on small $\xi\neq 0$
the  monopoles
and dyons become confined by strings. In fact, the elementary monopoles/dyons are represented by junctions
of two different elementary non-Abelian strings \cite{T,SYmon,HT2}, see also a detailed
discussion of the monopole/dyon confinement in Sect.~\ref{conf}. This
means that, as we move from the domain II into  III at small nonvanishing $\xi$ the
$W$ boson ``decays" into an antimonopole and dyon; however,  these states cannot 
abandon each other and move far apart because they
are confined. Therefore, the
$W$ boson evolves into a stringy meson formed by an antimonopole and dyon connected by
two strings, as shown in  Fig.~\ref{figmeson}, see \cite{SYrev} for a discussion of these
stringy mesons.

\begin{figure}
\epsfxsize=6cm
\centerline{\epsfbox{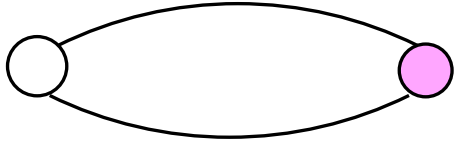}}
\caption{\small Meson formed by antimonopole and dyon connected by two strings.
Open and closed circles denote dyon and antimonopole respectively.}
\label{figmeson}
\end{figure}

These stringy mesons
have nonvanishing U(1) global charges with respect to the Cartan generators of the SU(3) subgroup
of the global group (\ref{c+f}) (above we discussed  only one $W$ boson of this type, related to the
$T^3$ generator, however, in fact, we have six different charged gauge boson/quark states of this type). 
In the equal
mass limit these globally charged stringy mesons combine with neutral (with respect to the group
(\ref{brokengroup})) stringy mesons formed by pairs of monopoles and antimonopoles (or
dyons and antidyons) connected by two strings, to
form the octet representation  of the SU(3) subgroup
of the global group (\ref{c+f}) (in general, the adjoint representation of SU$(N)$). 
They are heavy  in the domain III, with mass of the order of $\Lambda$.

We propose to identify these stringy mesons with $(N^2-1)$ adjoints 
 $M_P^{P'}$ ($P,P'=1,...,N$)
 of the SU$(N)$ subgroup with which we have already had an encounter
{\em en route} from the domain I to  III along the line $\Delta m_{AB}=0$, see Sect.~\ref{LEbulkdual}.

The same applies to the $q^{kK}$ quarks  ($K=N+1,...,N_f$)
of the domains I and II. As we go through the crossover
into the domain III at small $\xi$ $q^{kK}$ quarks evolve into stringy mesons formed by pairs of antimonopoles
and dyons connected by two strings, see Fig.~\ref{figmeson}. However, these states are unstable.
To see that this is indeed the case, please, observe
 that in the equal mass limit
these stringy mesons  fill the bifundamental representations $(N,\bar{\tN})$ and $(\bar{N},\tN)$ 
of the global group (\ref{c+fd}); hence, can decay into light dyons/dual gauge bosons with the same
quantum numbers.

To summarize, in the domain III we have  the dyons and dual gauge fields in the 
$(1,1)$, $(1,\tN^2-1)$, $(N,\bar{\tN})$ and $(\bar{N},\tN)$ representations of the 
global group (\ref{c+fd}).
They are light (with masses $\sim \tilde{g}\sqrt{\xi}$) and enter the low-energy
effective action (\ref{SIII}). In addition to these, we have stable neutral heavy (with masses $\sim\Lambda$)
stringy mesonic  $M$ fields formed by pairs of (anti)monopoles and dyons connected by two
strings, see Fig.~\ref{figmeson}. The set of stable states of this type  forms the  $(N^2-1,1)$ 
representation of (\ref{c+fd}).

In the domain  I a reversed situation takes place:
we have the quarks and gauge bosons in the 
$(1,1)$, $(N^2-1,1)$, $(\bar{N},\tN)$ and $(N,\bar{\tN})$ representations 
of the global group (\ref{c+f}).
They have masses of the order of $g\sqrt{\xi}$. In addition to these  ``elementary" states,
we also have stable neutral
stringy mesonic  $M$ fields in the $(1,\tN^2-1)$ representation
of (\ref{c+f}). The latter mesons are heavier, they have masses of the order of $\sqrt{\xi}$, due
to the presence of strings connecting the monopoles/dyons.

All other stringy mesons of the matrix $M_{A}^B$ are metastable and decay into elementary
excitations with the same global quantum numbers.

It is seen that non-Abelian confinement works in our theory as follows. It is a 
combined effect of the Higgs screening, ``decay" process on CMS and confining strings 
formation.
Strings always 
confine monopoles or dyons in both original and dual theories. These confined dyons have charges whose difference from 
the monopole charge can be screened in the given regime, for example, in the theory with
$N=3$, $\tN=2$ the dyon charge
$(0,0;\,1,1;\,0,0)$ in the domain II and 
$(0,0;\,1,2;\,0,0)$ in the domain III for the $T^3$ monopole $(0,0;\, 0, 1;\,0,0)$. 
 As we pass from the domain
I to  III, the screened quarks and gauge bosons ``decay" into (anti)monopoles and dyons  which
are still bound together in pairs by strings and form mesons. And vice versa, when we go from the domain
III to  I, the screened dyons and dual gauge fields of the dual theory 
(\ref{SIII}) ``decay" into pairs of confined
(anti)monopoles and dyons and form the corresponding stringy mesons.
In other words,    in both domains related by duality, I and III, the  elementary excitations in the given region
evolve into stringy composite mesons in the dual region and vice versa.

It is worth mentioning the $N_f=N$ theory 
studied in 
\cite{SYcross} as an important particular application of this picture. 
In this case the dual theory in the domain III is the Abelian U(1)$^N$
gauge theory. It has $N$ light Abelian dyons and photons. In addition to these states, it has
$(N^2-1)$  heavy neutral mesonic $M_{P}^{P'}$fields  which form the adjoint multiplet of the global
SU$(N)_{C+F}$ group. These states were identified in \cite{SYcross}. Here we reveal their
physical nature. They are  mesonic states formed by monopole/dyon pairs connected 
by two strings as shown in Fig.~\ref{figmeson}.

\section{World-sheet duality}
\label{2Ddual}
\setcounter{equation}{0}

In the previous sections we demonstrated that, as we reduce $\xi$ below $\Lambda^2$
and enter the domain III in Fig.~\ref{figphasediag},  our original microscopic U$(N)$ gauge theory with
$N_f$ flavors undergoes the crossover transition to the U$(\tN)\times U(1)^{N-\tN}$ gauge theory with
$N_f$ flavors. Now we show how this bulk duality is translated in the language of
the world-sheet duality on the non-Abelian string.

\subsection{Dual world-sheet theory}
\label{dwsth}

As was discussed in Sect.~\ref{strings}, if the quark mass differences are small, the
$(1 , ... ,\, N)$ vacuum of the original microscopic U$(N)$ gauge theory  supports non-Abelian 
semilocal strings. Their internal dynamics is 
described by the effective two-dimensional low-energy \ntwot sigma model (\ref{wcp}).
The model has $N$ orientational moduli $n^P$ with the U(1) charge $+1$ and masses $m_P=\{m_1,...,m_N\}$,
plus
$\tN$ size moduli $\rho^K$, with the U(1) charge $-1$ and masses $(-m_K)=-\{m_{N+1},...,m_{N_f}\}$.

Clearly, the dual bulk U$(\tN)$ theory (\ref{SIII}) in the domain III also supports
 non-Abelian semilocal strings. We found that the  $(1, ... ,\, N)$ vacuum  of the original theory transforms into the 
the
$(N+1,...,N_f,\tN+1,...,N)$ vacuum  of the dual theory. Therefore, the internal string dynamics on the string
world sheet
is described by a similar \ntwot sigma model. Now it has $\tN$ orientational moduli  with the 
U(1) charge $+1$ and masses $m_K=\{m_{N+1},...,m_{N_f}\}$. 
To make contact with (\ref{wcp}) let us call them $\trho^K$.
In addition,  it has $N$ size moduli  with the U(1) charge $-1$ and masses $(-m_P)=-\{m_1,...,m_N \}$. We 
refer to these  size moduli as 
$\tn^P$.

The bosonic part of the action of the world-sheet model 
in the gauge formulation (which assumes taking the limit $\tilde{e}^2\to\infty$) has the form
\beqn
S_{{\rm dual}} &=& \int d^2 x \left\{
 |\nabla_{\alpha} \trho^{K}|^2 +|\tilde{\nabla}_{\alpha} \tn^P|^2
 +\frac1{4e^2}F^2_{\alpha\beta} + \frac1{e^2}\,
|\pt_{\alpha}\sigma|^2
\right.
\nonumber\\[3mm]
&+&\left.
2\left|\sigma+\frac{m_P}{\sqrt{2}}\right|^2 \left|\tn^{P}\right|^2 
+ 2\left|\sigma+\frac{m_{K}}{\sqrt{2}}\right|^2\left|\trho^K\right|^2
+ \frac{e^2}{2} \left(|\trho^K|^2-|\tn^{P}|^2 -2\tilde{\beta}\right)^2
\right\},
\nonumber\\[3mm]
&& 
P=1,...,N\,,\qquad K=N+1,...,N_f\,,
\label{dcp}
\eeqn
where
\beq
\nabla_{\alpha}=\pt_{\alpha}-iA_{\alpha},\qquad \tilde{\nabla}_{\alpha}=\pt_{\alpha}+iA_{\alpha}.
\eeq
We see that the roles of orientational
and size moduli are interchanged in Eq.~(\ref{dcp})  compared with (\ref{wcp}).
As in the  model (\ref{wcp}), small mass differences
$(m_A-m_B)$  lift orientational and size zero modes of the non-Abelian semilocal string generating a
 shallow potential on the moduli space.
Much in the same way as in the model (\ref{wcp}), the dual coupling constant $\tilde{\beta}$ is
determined by the bulk dual coupling $\tilde{g}_2^2$,
\beq
4\pi\tilde{\beta}(\xi) = \frac{8\pi^2}{\tilde{g}_2^2}(\xi)=
(N-\tN )\ln{\frac{\Lambda}{\tilde{g}\sqrt{\xi}}}\gg 1 \,,
\label{d2coupling}
\eeq
see Eqs.~(\ref{betag}) and  (\ref{2coupling}). Here we take into account 
the fact that both the bulk and world-sheet
dual theories  have identical $\beta$ 
functions, with the first coefficient $(\tN-N)<0$. They are both IR free;
 therefore, the coupling constant $\tilde{\beta}$ is positive at 
$\Lambda \gg \sqrt{\xi}$. As in the model (\ref{wcp}), the coincidence of $\beta$ functions of the bulk
and world-sheet theories implies that the scale of the dual model (\ref{dcp}) is equal to that of the 
bulk theory,
$$\tilde{\Lambda}_{\sigma}=\Lambda\,,$$ cf. (\ref{lambdasig}).
Comparing (\ref{d2coupling}) with (\ref{2coupling}) we see that 
\beq
\tilde{\beta}=-\beta.
\label{tbb}
\eeq
Thus, the dual theory (\ref{dcp}) can be interpreted as a continuation of the sigma model (\ref{wcp})
to  negative values of the coupling constant $\beta$.

Note also, that both dual world-sheet theories (\ref{wcp}) and (\ref{dcp}) give {\em effective low-energy}
 descriptions of  string dynamics and are applicable only at scales well below $g\sqrt{\xi}$.

Concluding this section a comment is in order
regarding  the world-sheet duality between
two-dimensional sigma models (\ref{wcp}) and (\ref{dcp}). It was previously noted in Ref.~\cite{Jsem}. 
In this paper two bulk theories,  with the U$(N)$ and U$(\tN)$ gauge groups, were considered
(these theories were referred to as a dual pair in \cite{Jsem}). Two-dimensional 
sigma  models (\ref{wcp}) and (\ref{dcp})
were presented as  effective low-energy descriptions of the non-Abelian strings for 
these two bulk theories.

\subsection{The BPS spectrum}
\label{bpsspec}

Dorey noted \cite{Dorey} that the exact BPS spectrum  of two-dimensional \ntwot
$CP(N-1)$ model (\ref{cpg}) coincides with the BPS spectrum of massive states in four-dimensional
\ntwo QCD (\ref{model}) with the U(N) gauge group and $N_f=N$ flavors in the $r=N$ vacuum on the Coulomb branch
 (i.e. at $\xi=0$). Later, this correspondence of the BPS spectra was generalized to cover  the $N_f>N$ case \cite{DoHoTo}.
Namely, it was shown that the BPS spectrum of kinks in the two-dimensional model (\ref{wcp}) coincides
with the BPS spectrum of massive monopoles and dyons in the $r=N$ vacuum on the Coulomb branch of the 
four-dimensional theory (\ref{model}). 

The reason for this amazing coincidence was understood and explained later in Ref.~\cite{SYmon,HT2},  for a review
see \cite{SYrev}. Consider the 
bulk theory (\ref{model}) at large $\xi$.
As was discussed above,
it is the  monopoles that are confined by strings. Elementary monopoles are represented by string junctions of two different elementary non-Abelian strings \cite{T,SYmon,HT2}.
 Each string of the bulk theory corresponds to a particular vacuum of the world-sheet theory. In particular,
the \ntwot supersymmetric sigma model (\ref{wcp}) on the string 
has $N$ degenerate vacua and kinks interpolating between distinct vacua. These kinks are interpreted as 
confined monopoles of the bulk theory \cite{T,SYmon,HT2}.

Please observe that the mass of the confined BPS monopole (a.k.a sigma-model kink) is a holomorphic function
on the parameter space and, therefore, cannot depend  \cite{SYmon,HT2} on the nonholomorphic parameter $\xi$.
Thus we can reduce $\xi$ all the way to $\xi=0$ and the mass of the confined monopole stays intact.
At $\xi=0$, on the Coulomb branch, the monopoles are no longer confined and their masses are given
by the exact Seiberg--Witten solution of the bulk theory. This leads us to the conclusion that the kink masses 
in the two-dimensional sigma model (\ref{wcp}) should coincide with those of monopoles/dyons 
in the four-dimensional bulk theory on the Coulomb branch in the $r=N$ vacuum. As was mentioned above, this
fact was earlier observed``experimentally"  in \cite{Dorey,DoHoTo}.

Now the same logic leads us to one another conclusion. Since the confined monopole masses in the bulk theory
 do not depend on $\xi$,  we can reduce $\xi$ and safely pass from the domain I to III, keeping 
 the BPS spectrum unchanged. In the domain I the spectrum of confined monopoles is
 determined by the  BPS spectrum of the sigma model (\ref{wcp}),
while in the domain III it is determined by the BPS spectrum of the dual sigma model (\ref{dcp}).
Thus, we arrive at the conclusion, that BPS spectra of two dual world-sheet models (\ref{wcp}) and (\ref{dcp}) 
should coincide.

It is instructive to  explicitly check this assertion. Let us
 start from (\ref{wcp}) at positive $\beta$
and  use the description
of the supersymmetric  model (\ref{wcp}) in terms of exact superpotentials \cite{HaHo,DoHoTo}.
 Following \cite{W93} and integrating out fields $n^P$ and $\rho^K$  we can describe
 the model   by an
exact twisted superpotential 
\beqn
 {\cal W}_{\rm eff}
 & =& 
\frac1{4\pi}\sum_{P=1}^N\,
\left(\sqrt{2}\,\Sigma+{m}_P\right)
\,\ln{\frac{\sqrt{2}\,\Sigma+{m}_P}{\Lambda}}
\nonumber\\[3mm]
&-& 
\frac1{4\pi}\sum_{K=N+1}^{N_F}\,
\left(\sqrt{2}\,\Sigma+{m}_K\right)
\,\ln{\frac{\sqrt{2}\,\Sigma+{m}_K}{\Lambda}} 
\nonumber\\[3mm]
&-& \frac{N-\tN}{4\pi} \,\sqrt{2}\,\Sigma\, ,
\label{2Dsup}
\eeqn
where $\Sigma$ is a twisted superfield \cite{W93} (with $\sigma$ being its lowest scalar
component). 
Minimizing this superpotential with 
respect to $\sigma$ we find
\beq
\prod_{P=1}^N(\sqrt{2}\,\sigma+{m}_P)
=\Lambda_{\sigma}^{(N-\tN)}\,\prod_{K=N+1}^{N_f}(\sqrt{2}\,\sigma+{m}_K)\,.
\label{sigmaeq}
\eeq
Note that the roots of this equation coincide with the double roots of the Seiberg--Witten curve (\ref{curve}) of 
the bulk theory \cite{Dorey,DoHoTo}.

The BPS kink masses are given by differences of the superpotential (\ref{2Dsup}) calculated at distinct roots,
\beq
m^{\rm BPS}_{ij}=2\left|{\cal W}_{\rm eff}(\sigma_i)-{\cal W}_{\rm eff}(\sigma_j)\right|.
\label{BPSmass}
\eeq
It is easy to show that above masses coincide with those of monopoles and dyons in the bulk theory 
given by the period integrals of the Seiberg--Witten curve presented in (\ref{periodint}) (this equation
 is written down for 
certain particular roots). As was mentioned above, this coincidence of
the  BPS spectra of the
world-sheet  and bulk theories was expected.

Now let us consider the effective superpotential of the dual world-sheet theory (\ref{dcp}).
It has the form
\beqn
 {\cal W}^{\rm dual}_{\rm eff} 
 &= &
\frac1{4\pi}\sum_{K=N+1}^{N_f}\,
\left(\sqrt{2}\,\Sigma+{m}_K\right)
\,\ln{\frac{\sqrt{2}\,\Sigma+{m}_K}{\Lambda}}
\nonumber\\[3mm]
& -& 
\frac1{4\pi}\sum_{P=1}^{N}\,
\left(\sqrt{2}\,\Sigma+{m}_P\right)
\,\ln{\frac{\sqrt{2}\,\Sigma+{m}_K}{\Lambda}}
\nonumber\\[3mm]
 &-& \frac{\tN-N}{4\pi} \,\sqrt{2}\,\Sigma \, .
\label{2Ddsup}
\eeqn
We see that it coincides with the superpotential (\ref{2Dsup})  up to the sign. Clearly, both the
root equations  and the BPS spectra  are the same for both dual sigma
models, as expected. They are given by Eqs.~(\ref{sigmaeq}) and (\ref{BPSmass}), respectively.

\section{Conclusions}
\label{conclu}

In this paper  we continued our explorations of the transition from weak to strong coupling
in \ntwo supersymmetric QCD in the course of variation of the parameter $\xi$. These explorations began in 
\cite{SYcross} where we analyzed the case $N_f=N$
and discovered a crossover transition from the original weakly coupled (at large $\xi$) non-Abelian theory
to a strong coupling regime (at small $\xi$) described by a dual weakly coupled Abelian theory.
Now we expanded this study to cover the $N_f>N$ case.

We found that at strong coupling (i.e. small $\xi$) a dual  weakly coupled ${\mathcal N}=2$
theory exists but it is non-Abelian,
based on the gauge group ${\rm U}(\tN)$.
This non-Abelian dual  describes low-energy physics at small $\xi$.
The dual theory has $N_f$ flavors
of light dyons, to be compared with $N_f$ quarks in the original U($N$)
theory. Both, the original and dual theories are Higgsed and share the same global symmetry
${\rm SU}(N)\times  {\rm SU}(\tN) \times {\rm U}(1)$, albeit
the physical meaning of the ${\rm SU}(N)$ and $ {\rm SU}(\tN)$ factors
is different in the large- and small-$\xi$ regimes.
Both regimes support non-Abelian semilocal strings.

 In each of these two regimes
particles that are in the adjoint representations with respect to one of the factor groups
exist in two varieties: elementary fields and composite states bound by strings.
These varieties interchange upon transition from one regime to the other. 
We conjecture that the composite stringy states can be related to Seiberg's $M$ fields. 

We demonstrated that non-Abelian confinement in our theory  is a 
combined effect of the Higgs screening, ``decay" processes on CMS and confining string
formation.
Strings always 
confine monopoles or dyons (whose charges can be represented as a sum of a monopole 
and  $W$-boson charges)
in both original and dual theories. 
 As we pass from the domain
I to  III, the screened quarks and gauge bosons ``decay" into (anti)monopoles and dyons  which
are still bound together in pairs by strings and form mesons. 
These mesons form the adjoint representation of 
the ${\rm SU}(N)$ factor of the global group.
And vice versa, when we go from the domain
III to  I, the screened dyons and dual gauge fields of the dual theory 
 ``decay" into pairs of confined
(anti)monopoles and dyons and form the corresponding stringy mesons
which fall into the adjoint representation of the ${\rm SU}(\tN)$ factor of the global group.
A level crossing takes place on the way.

 The bulk duality that we observed translates into a two-dimensional
duality on the world sheet of the non-Abelian strings. At large $\xi$ the 
internal dynamics of the semilocal non-Abelian strings is described by
the sigma model of $N$ orientational and $(N_f-N)$ size moduli, while
at small $\xi$ the roles of orientational and size moduli interchange.
The BPS spectra of two dual sigma models (describing confined 
monopoles/dyons of the bulk theory) coincide.

 It would be extremely interesting to trace parallels
between the non-Abelian duality we detected and string theory constructions.
We conjecture that such parallels must exist.

\section*{Acknowledgments}
This work  is supported in part by DOE grant DE-FG02-94ER408. 
The work of A.Y. was  supported 
by  FTPI, University of Minnesota, 
by RFBR Grant No. 09-02-00457a 
and by Russian State Grant for 
Scientific Schools RSGSS-11242003.2.

\end{document}